\documentclass[a4paper,fleqn]{cas-sc}

\usepackage{lineno}
\usepackage[authoryear]{natbib}
\bibpunct{(}{)}{;}{a}{,}{,}

\def\tsc#1{\csdef{#1}{\textsc{\lowercase{#1}}\xspace}}
\tsc{WGM}
\tsc{QE}
\tsc{EP}
\tsc{PMS}
\tsc{BEC}
\tsc{DE}
\usepackage{rotating}
\usepackage{multirow}
\usepackage{lscape}
\usepackage{graphicx}
\usepackage{xcolor}
\usepackage{arydshln}
\usepackage{amssymb}
\usepackage[shortlabels]{enumitem}
\usepackage{float}

\begin{document}
\let\WriteBookmarks\relax
\def\floatpagepagefraction{1}
\def\textpagefraction{.001}
\shorttitle{Composition of L5 Martian Trojans from spectroscopy}
\shortauthors{A.~A.~Christou et~al.}

\title [mode = title]{Composition and origin of L5 Trojan asteroids of Mars: Insights from spectroscopy}                      

\author[1]{Apostolos A.~Christou}
\cormark[1]
%\fnmark[1]
%\ead{Apostolos.Christou@Armagh.ac.uk}
\address[1]{Armagh Observatory and Planetarium, College Hill, BT61 9DG Armagh, United Kingdom}

\author[1,2]{Galin Borisov}[orcid=0000-0002-4516-459X]
%\cormark[1]
%\fnmark[2]
%\ead{Galin.Borisov@Armagh.ac.uk}
\address[2]{Institute of Astronomy and NAO, 72 Tsarigradsko Chauss\'ee Blvd., BG-1784 Sofia, Bulgaria}

\author[3]{Aldo Dell'Oro}
\address[3]{INAF - Osservatorio Astrofisico di Arcetri, Largo E. Fermi 5, I-50125, Firenze, Italy}

\author[4]{Alberto Cellino}
\address[4]{INAF - Osservatorio Astrofisico di Torino, via Osservatorio 20, 10025 Pino Torinese, Italy}

\author[5]{Maxime Devog\`{e}le}[orcid=0000-0002-6509-6360]
\address[5]{Lowell Observatory, 1400 W Mars Hill RD, Flagstaff, AZ 86001, USA}

%\cortext[cor1]{Corresponding author}

\begin{abstract}
    We investigate the mineralogical makeup of L5 Martian Trojan asteroids via reflectance spectroscopy, paying special attention to (101429) 1998 $\mbox{VF}_{31}$, the only L5 Trojan that does not belong to the Eureka family \citep{Christou2013}. We find that this asteroid most likely belongs to the Bus-Demeo S-complex, in agreement with \citet{Rivkin2007}. We compare it with a variety of solar system bodies and obtain good spectral matches with Sq- or S-type asteroids, with spectra of the lunar surface and of Martian and lunar meteorites. Mixture fitting to spectral endmembers suggests a surface abundance of Mg-rich orthopyroxene and iron metal or, alternatively, a combination of plagioclase and metal with a small amount of Mg-poor orthopyroxene. The metallic component may be part of the intrinsic mineral makeup of the asteroid or an indication of extreme space weathering. 
    
    In light of our findings, we discuss a number of origin scenarios for (101429). The asteroid could be genetically related to iron-rich primitive achondrite meteorites (\citeauthor{Rivkin2007}), may have originated as impact ejecta from Mars - a scenario proposed recently for the Eureka family asteroids \citep{Polishook_Nature} - or could represent a relic fragment of the Moon's original solid crust, a possibility raised by the asteroid's close spectral similarity to areas of the lunar surface. If, on the other hand, (101429) is a relatively recent addition to the Martian Trojan clouds \citep{Christou2020}, its origin is probably traced to high-inclination asteroid families in the Inner Main Belt.
    
    For the olivine-dominated Eureka family, we find that the two smaller asteroids in our sample are more spectrally similar to one another than to (5261) Eureka, the largest family member. Spectral profiles of these three asteroids are closely similar shortward of $\sim$0.7 $\mu$m but diverge at longer wavelengths. For the two smaller asteroids in particular, we find the spectra are virtually identical in the visible region and up to $0.8$ $\mu$m. We attribute spectral differences in the near-IR region to differences in either: degree of space weathering, olivine chemical composition and/or regolith grain size. 
\end{abstract}

%\begin{highlights}
%\item We study reflectance spectra of four Mars-trailing Trojans as %diagnostics of surface composition  
%\item One asteroid has different composition than the others, being more pyroxene- and iron-rich 
%\item It likely originated from Mars, a large differentiated asteroid or the Moon. 
%\end{highlights}


\maketitle

\section{Introduction}
\let\thefootnote\relax\footnotetext{* Corresponding author.\\ \indent \indent \indent
{\it E-mail addresses:} \href{mailto:Apostolos.Christou@Armagh.ac.uk}{Apostolos.Christou@Armagh.ac.uk} (A.A. Christou), \href{mailto:Galin.Borisov@Armagh.ac.uk}{Galin.Borisov@Armagh.ac.uk} (G. Borisov).}

\let\thefootnote\relax\footnotetext{\textcopyright 2020 Elsevier Inc. All rights reserved. This manuscript version is made available under the CC-BY-NC-ND 4.0 license \\ \indent \indent \indent 
\href{http://creativecommons.org/licenses/by-nc-nd/4.0/}{http://creativecommons.org/licenses/by-nc-nd/4.0/}}

Trojan asteroids orbit the Sun near the L4 and L5 Lagrangian equilibrium points $60^{\circ}$ ahead or behind a planet along its orbit \citep{Dermott_SSD}. Only three planets in the solar system, namely Mars, Jupiter and Neptune, are attended by dynamically stable Trojans. Because of their orbital stability, these objects allow us to constrain models of the earliest stages of formation of our solar system. The few known Trojans of Mars are the only known stable population of asteroids in the terrestrial planet region. They are most likely material anchored to the Martian orbit just as the solar system architecture was reaching its final configuration \citep{Scholl2005}.

\citet{Christou2013} found that all Mars trailing Trojans, except one, form an orbital family with (5261) Eureka as its largest member. Family members show a common, olivine-dominated composition \citep{Rivkin2007,Lim2011,Borisov2017,Polishook_Nature} that is generally rare among asteroids \citep{Sanchez2014,DeMeo2019}.  
The only other known L5 Trojan is (101429) 1998 $\mbox{VF}_{31}$. Compositional information for this object is limited to single visible and near-infrared reflectance spectra \citep{Rivkin2003,Rivkin2007} as well as an albedo/size determination from thermal infrared data \citep{Trilling2007}. These indicate a similarity to S-type asteroids, quite different from the olivine-dominated Trojans and with a high abundance of pyroxene.

Here we take a fresh look at the L5 Trojans using new observations of (101429) and a more detailed examination of recently available spectra of Eureka family asteroids. Our ultimate aim is to obtain insight on the Trojans' original parent bodies and help incorporate them into narratives of the solar system's early evolution. Our study benefits from the availability of new models \citep{Andrews-Hanna_Natur,Polishook_Nature,Neumann-2018_Icarus} and of spectral databases and tools for comparing asteroids to astrophysically relevant materials and surfaces \cite[eg][]{DeMeoDB,M4AST}. 
\begin{table*}[pos=h]
\centering
\caption[Observational circumstances for observations used in this work.]
{Observational circumstances for observations used in this work.}
\label{phaseang}
\begin{tabular}{lcllr}
\hline 
\hline      
Asteroid${}^{\dagger}$  & Date & Instrument  &  Solar analogue${}^{\ddagger}$  &  Phase angle\\
\hline
	(101429) 1998 $\mbox{VF}_{31}$   & 27 Jan 2016 & XSHOOTER   & HD\,44594             & 41.5$^\circ$\\
	(101429) 1998 $\mbox{VF}_{31}$   & 11 May 2005 & SpeX       & L102-1081, L110-361 & 26.2$^\circ$\\
	(385250) 2001 $\mbox{DH}_{47}$   & 02 Feb 2016 & XSHOOTER   & HD\,67010             & 12.5$^\circ$\\
	(311999) 2007 $\mbox{NS}_{2}$    & 02 Mar 2016 & XSHOOTER   & HD\,67010             & 27.3$^\circ$\\
	(5261) Eureka & 19 May 2005 & SpeX       & L102-1081, L105-56  & 4.5$^\circ$\\
\hline
\multicolumn{5}{l}{\parbox{122mm}{${}^{\ddagger}$ Standard stars from \citep{Landolt} and spectral solar analogues from \citep{SSA}}}
\end{tabular}
\end{table*}

The paper is organised as follows: in the next Section we describe the observational data obtained for this work and the methods used to spectrally compare the asteroids to other solar system bodies and laboratory-measured materials. In Section~3 we focus on (101429) where we taxonomically rank the asteroid spectrum and search for surfaces and materials that best match its spectral characteristics. Section~4 shifts the focus of the paper to the Eureka family asteroids, highlighting spectral differences and attempting to explain them. Finally, Section~5 presents our conclusions, discusses the implications of our findings and charts out avenues for future work.

\section{Methods}
\subsection{Observations and data reduction}
Observations used in this work are summarised in Table~\ref{phaseang}. All the asteroids except Eureka were observed with X-SHOOTER \citep{Veretal11} at the ESO Very Large Telescope facility. Details of the data reduction for the two Eureka family members discussed in Section~4 may be found in \citet{Borisov2017}. Data for (101429) 1998 $\mbox{VF}_{31}$, VF31 hereafter, was likewise obtained over the spectral range 300-2480\,nm through the ultraviolet \ blue, visible (VIS), and near-infrared (NIR) arms of the instrument. Observations were carried out in nodding mode and reduced with the ESO Reflex pipeline version 2.6.8 \citep{ESOReflex}. The spectrum was then smoothed using a running average with wavelength steps of 5\,nm in the visible and 30\,nm in the NIR regions. The UVB part of the obtained spectrum was discarded because of its low {\it S/N}. Finally, the binned asteroid spectrum was divided by the spectrum of the solar analogue star, which was reduced in the same way and observed on the same night, and the obtained instrumental reflectance was normalised to unity at 550\,nm. The data for Eureka and the NIR part of the VF31 spectrum - kindly provided by A.~Rivkin - were obtained using the SpeX instrument in prism mode on the Infrared Telescope Facility (IRTF) on Mauna Kea, Hawaii. Details on data reduction, spectra extraction and wavelength calibration can be found in \citet{Rivkin2007} and references therein. The Rivkin NIR spectrum of VF31 was smoothed using a running average with wavelength step of 20\,nm and then resampled to the same wavelength locations as our spectrum.

\begin{figure*}[pos=h!]
\includegraphics[width=0.7\columnwidth]{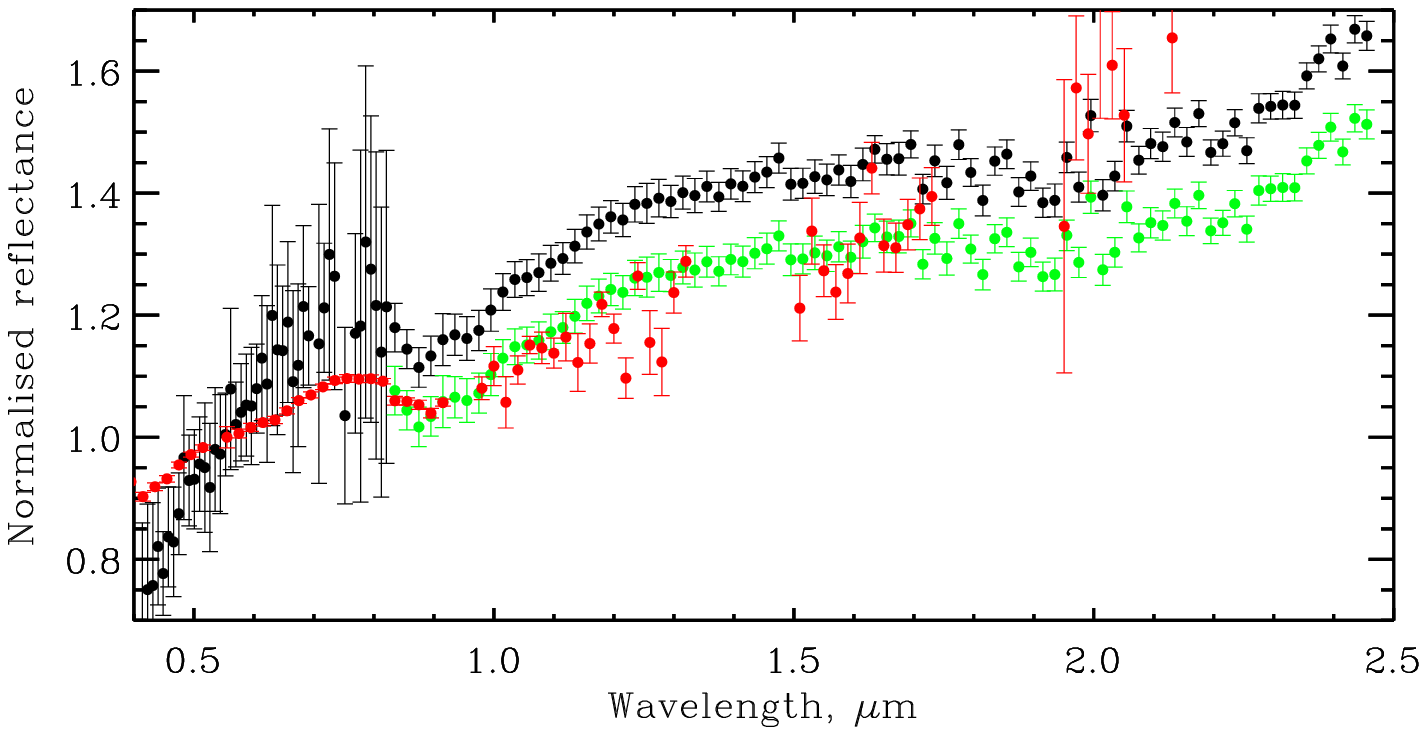} 
\includegraphics[width=0.7\columnwidth]{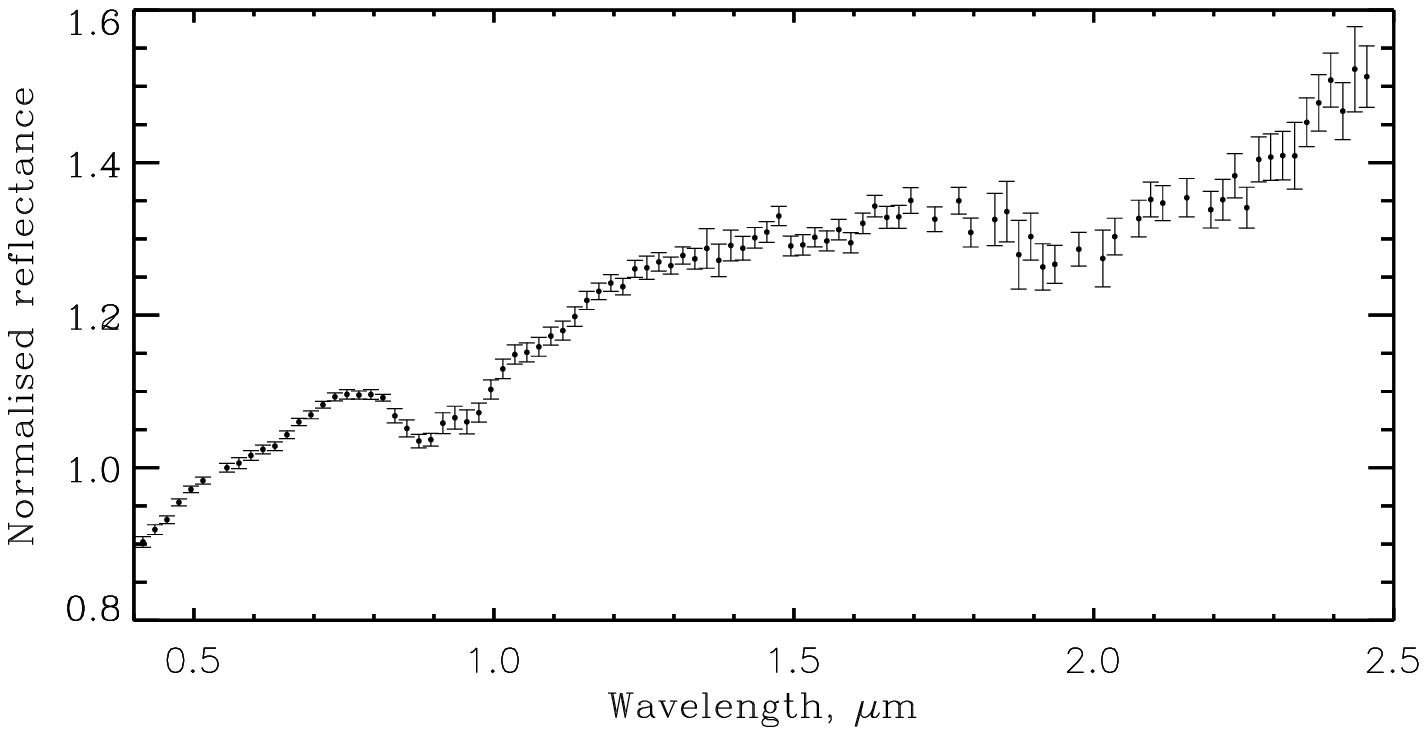} 
\caption{{\it Top:} The \citet{Rivkin2007} spectrum of VF31 (black), our VLT spectrum (red) and the rescaled near-infrared \citeauthor{Rivkin2007} spectrum (green). {\it Bottom:} The composite VF31 spectrum used in this work.}
\label{VisIRR}
\end{figure*}

To produce a visible and near-infrared (NIR) spectrum of VF31 suitable for analysis, we combine the NIR spectrum from \citet{Rivkin2007} with our X-SHOOTER visible spectrum (Figure~\ref{VisIRR}, top panel) to take advantage of the higher S/N in our visible spectrum and in \citeauthor{Rivkin2007}'s NIR spectrum. We do this by rescaling Rivkin's NIR spectrum to match our visible and NIR data in the overlapping wavelength interval below 1.8\,$\mu$m, using the function
\begin{equation}
\chi^2=\sum\limits_{i}^{N}{\frac{(R_i - f \times R^{\rm Rvkn}_i)^2}{\sigma_i^2}}
\label{scalingfactor}
\end{equation}
where $R_i$ and $R^{\rm Rvkn}_i$ are ours and Rivkin's reflectance measurements at the $i^{\rm th}$ wavelength position respectively,\newline \mbox{$f\in[0.80, 1.05]$} is the scaling factor and $\sigma_i$ are the measurement uncertainties. 
The minimum $\chi^2$ value is achieved for a scaling factor of $f=0.9125$, which we adopt here. The resulting composite spectrum we use for taxonomic classification and search for spectral analogues in the following Sections is presented in the bottom panel of Figure~\ref{VisIRR}.

\subsection{Comparing spectra using cost functions}\label{match}
In the following Sections we compare the VF31 spectrum with other spectra to find the best matches. The matching procedure is based on defining suitable cost functions. If the observed spectrum i.e.~the spectrum of VF31 is composed of $N$ measurements $x_i$ at the corresponding wavelengths $\lambda_i$, and assuming that the corresponding values $\mu_i$ of the model spectrum (i.e.~other spectra) are known, the cost function measures the deviation between the two spectra. Here we use the following curve-matching cost functions \cite[ Eqs~5 and 6 resp.~from][]{M4AST}:
\begin{equation}
\Phi_{\rm std}={1 \over N}\sqrt{\sum\limits_{i}^{N}{(e_i-\bar{e})^2}},\chi^2_{\mu}=\sum\limits_i^N{e_{i}^2 \over \mu_i},
\label{phi}
\end{equation}
where $e_i=x_i-\mu_i$ and $\bar{e}$ the mean value over all measurements, $x_i$ is the 
measurement at the $i^{\rm th}$wavelength position, and $\mu_i$ is the model evaluation at the same location.
These are computationally straightforward to implement for each model spectrum and neither function depends on the measurement uncertainties. At the same time, the use of two different cost functions instead of one improves the robustness of the procedure. Note that the second of our statistical functions is slightly modified from Eq.~6 in \citeauthor{M4AST} \cite[originally from][]{NEDELCU07} in that we divide by $\mu_{i}$ rather than $x_{i}$. For this reason, we introduce the ``$\mu$'' subscript to distinguish between our version of the statistic and that in \citeauthor{M4AST}.

Given an ensemble of model spectra interpolated at the same $N$ wavelengths $\lambda_i$ as the observed spectrum, the spectra most similar to the observed spectrum are those for which the value of the cost function is lowest. In practical terms we can build a score list, one for $\chi^2_\mu$ and one for $\Phi_{\rm std}$, where the models are ordered by increasing value of the cost function. On the other hand, each value $x_i$ is affected by a measurement error $\sigma_i$, and each measurement $x_i$ is only one of the possible values coming from a statistical distribution characteristic of the measurement process. So a model could be placed at the top of the score list only by chance. Starting from some statistical assumptions about the fluctuations of the values of the observed spectrum (see Appendix~\ref{sec:app1}), we estimate the probability $P$ that a particular model spectrum will be at the top of the score list.

\begin{table}[pos=h]
\centering
\caption[Taxonomic ranking of the composite spectrum of VF31.]
{ Taxonomic ranking of the composite spectrum of VF31.}
\label{taxrankVNIR-SR}
\begin{tabular}{lr|lr}
\hline 
\hline      
\multicolumn{1}{c}{Taxonomic} & $P_{\chi^2_{\mu}}$ &  \multicolumn{1}{c}{Taxonomic} & $P_{\Phi_{\rm std}}$  \\
\multicolumn{1}{c}{class} & \multicolumn{1}{c|}{\%} &  \multicolumn{1}{c}{class} & \multicolumn{1}{c}{\%}  \\
\hline
S     &    24.2  &    Xk    &    28.4  \\
Xn    &    17.4  &    Xn    &    24.0  \\
D     &    15.2  &    S     &    11.5  \\
Xk    &    12.7  &    Cgh   &     9.1 \\
L     &    12.5  &    Ch    &     5.9 \\
T     &     7.9  &    L     &     5.2 \\
\hline
\end{tabular}
\end{table}

\begin{figure}[pos=h!]
\centering
\includegraphics[width=0.3\columnwidth]{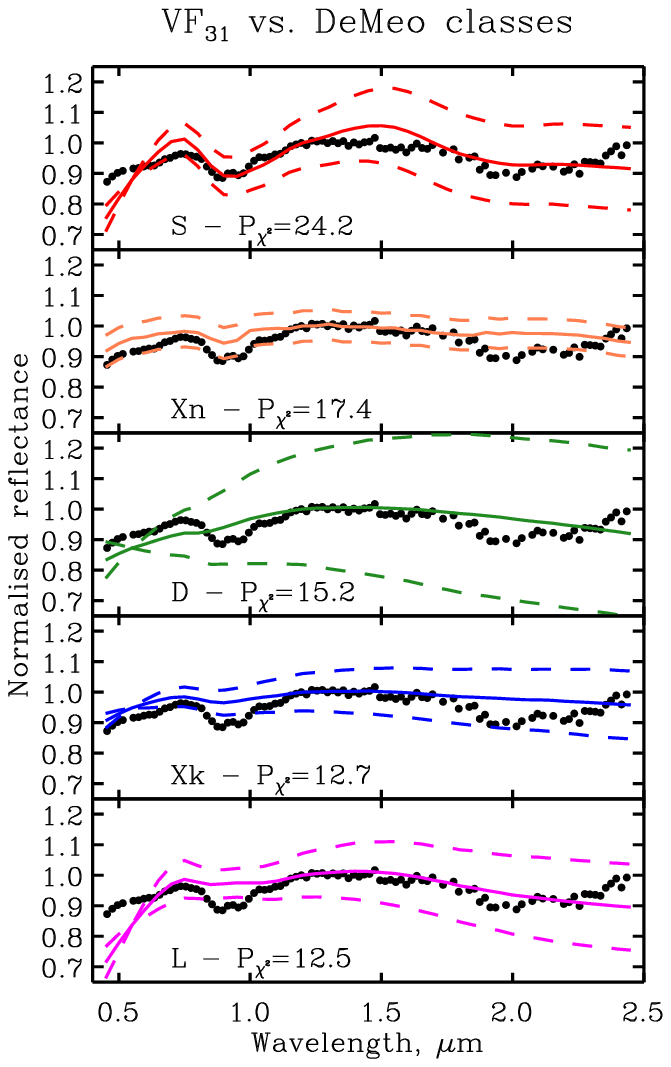}
\includegraphics[width=0.3\columnwidth]{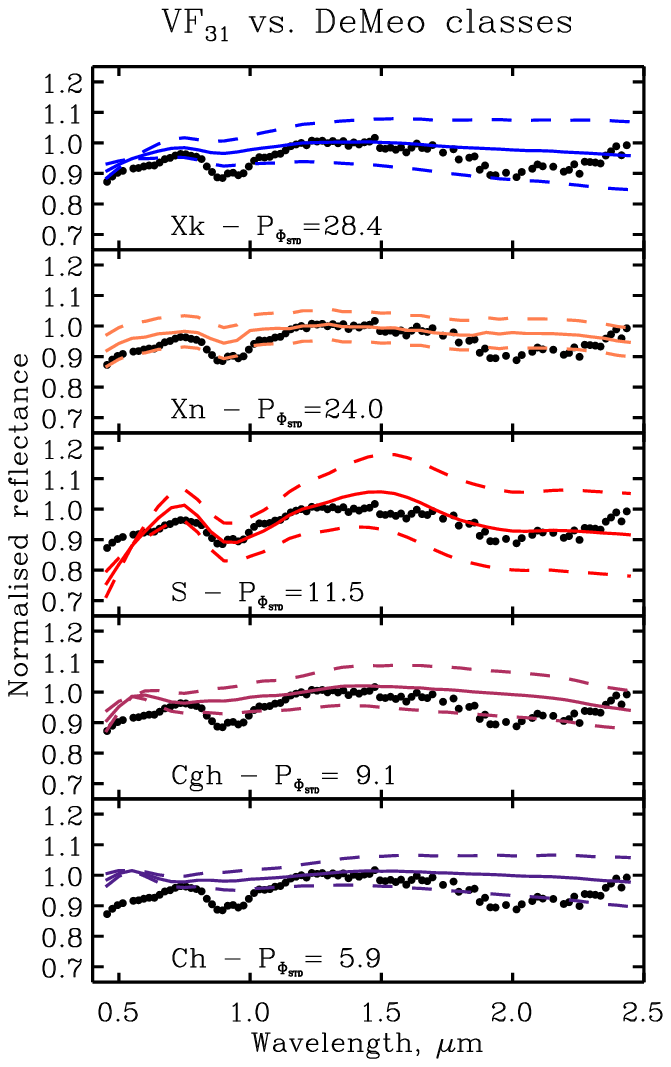}
\caption{Visible-NIR taxonomic ranking with DeMeo average classes using $\chi^2_{\mu}$ (left panel) and $\Phi_{\rm std}$ (right panel).}
\label{VNIR_TAX}
\end{figure}

\section{(101429) 1998 \mbox{VF}$_{31}$}
\subsection{Taxonomic classification}
\label{tax-visible}

To determine the taxonomy of VF31, we compare the observed spectrum with taxonomic class average spectra from \citet{DeMeoDB} using the ranking approach described in the previous Section. 

Both observed and model spectra are normalised at a common wavelength ($x_n = \mu_n = 1$ at the normalisation wavelength $\lambda_n$). We chose to normalise spectra at $\lambda_n=1.2$\,$\mu$m, because this position is away from absorption features at 1 and 2\,$\mu$m and free from strong telluric absorption. The same choice was adopted by \citet{DeMeo2019} in their investigation of NEA spectra when only the NIR part of the spectrum was available for analysis. Finally, we remove the spectral slope before performing the match, placing more emphasis on spectral shape variations and the locations and depths of absorption features. 

Table~\ref{taxrankVNIR-SR} lists taxonomies that scored better than $P_{\chi^2_\mu}=3\%$ (left) and the top five hits also for $P_{\Phi_{\rm std}}$ (right). Four out of the six taxonomic classes appear in both lists. Figure~\ref{VNIR_TAX} shows the highest-scoring taxonomies for both the ${\chi^2_\mu}$ and $\Phi_{\rm std}$ statistics. The lists include taxonomic end-members (D, L and T)  which, however, do not represent the VF31 spectrum well as they are devoid of strong spectral absorptions. The same is true for primitive taxonomies in the ${\Phi_{\rm std}}$ list (Cg and Cgh; not shown here).
Our feature-oriented approach favours the S and Xk taxonomic classes, although Xk lacks a 2\,$\mu$m absorption. S-types exhibit the same major absorption features as VF31 and, in this sense, are the best taxonomic match to the asteroid.
\begin{figure}[pos=htb]
\centering
\includegraphics{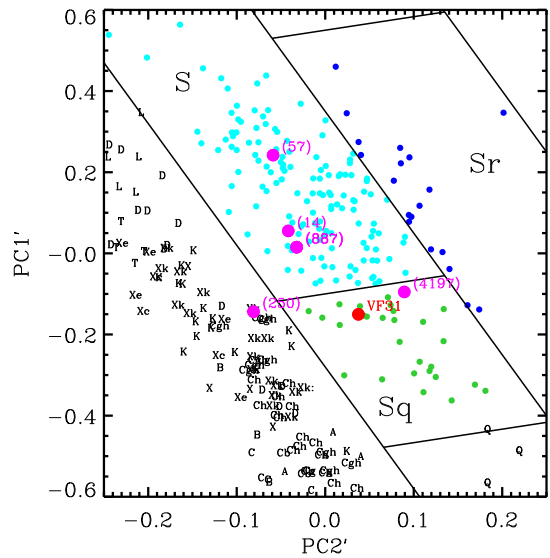} 
\caption{Location of VF31 (red point) in the PC2' vs PC1' principal component space compared to other asteroids. The 5 highest-ranked asteroids in the $P_{\chi^2_{\mu}}$-sorted list (Table~\ref{taxrankast-SR}) are shown in magenta. Non S-complex asteroids are represented by black characters indicating the taxonomic class.}
\label{PC1PC2}
\end{figure}

We complemented our ranking approach by processing the asteroid spectrum with the SMASS MIT Principal Component Analysis (PCA) software tool available online\footnote{\scriptsize http://smass.mit.edu/busdemeoclass.html}.
The s/w returns, apart from the slope, the five principal components PC1${}^{\prime}$ through PC5${}^{\prime}$ and the Bus-DeMeo taxonomic type \citep{DeMeoDB}. VF31 plots near the Sq/S boundary in PC2${}^{\prime}$ vs PC1${}^{\prime}$ principal component space (Fig.~\ref{PC1PC2}). Sq does not appear in our $P$-sorted list (Table~\ref{taxrankVNIR-SR}) but S is visually the most satisfactory fit in Figure~\ref{VNIR_TAX}, although its spectral profile variations are somewhat stronger than in the VF31 spectrum. 

The SMASS tool returns a slope value of 0.2292 $\mbox{$\mu$m}^{-1}$. We have reproduced this using our own fitting procedure and by normalising the spectrum at 0.55 $\mbox{$\mu$m}$, yielding a formal 1-sigma error of $\pm$0.0127 $\mbox{$\mu$m}^{-1}$. \citet{DeMeoDB} placed S-type spectra with slopes $>$0.25 $\mbox{$\mu$m}^{-1}$ into a separate class, Sw. VF31 most likely lies outside the domain of this class, provided that the composite spectrum used here to compute the slope does not suffer from strong systematic effects.

It appears that an S-complex taxonomy is the most likely match to VF31. In the next Section we investigate further by searching among candidate ``spectral analogue'' objects and surfaces.

\begin{table*}[pos=h]
\centering
\caption[Spectral comparison of VF31 with asteroids from the SMASS II and MITHNEOS surveys.]
{Spectral comparison of VF31 with asteroids from the SMASS II and MITHNEOS surveys.}
\label{taxrankast-SR}
\begin{tabular}{rllcr|rllcr}
\hline 
\hline      
\multicolumn{2}{c}{Asteroid} & Tax.  & Geom.  & \multirow{2}{*}{$P_{\chi^2_{\mu}}$} & \multicolumn{2}{c}{Asteroid} & Tax.  & Geom.  & \multirow{2}{*}{$P_{\Phi_{\rm std}}$} \\
No. & Name   & Class & Albedo &  &No. & Name   & Class & Albedo & \\
\hline\
(4197) & Morpheus   & Sq & 0.370 & 21.8 &  (14) & Irene      & S  & 0.159 & 17.6 \\
  (14) & Irene      & S  & 0.159 & 18.0 & (250) & Bettina    & Xk & 0.112 & 17.1 \\
 (250) & Bettina    & Xk & 0.112 & 14.1 &(4197) & Morpheus   & Sq & 0.370 & 16.1 \\
  (57) & Mnemosyne  & S  & 0.215 &  6.1 & (887) & Alinda     & S  & 0.310 &  7.6 \\
   (6) & Hebe       & S  & 0.268 &  4.6 &  (57) & Mnemosyne  & S  & 0.215 &  4.5 \\
 (887) & Alinda     & S  & 0.310 &  4.6 & (110) & Lydia      & X  & 0.181 &  4.1 \\
 (925) & Alphonsina & S  & 0.248 &  3.3 &   (6) & Hebe       & S  & 0.268 &  3.5  \\
\hline
\end{tabular}
\end{table*}

\subsection{Search for spectral analogues}
\subsubsection{Asteroids}
\label{asteroids}
\begin{figure*}
\includegraphics[width=0.33\textwidth]{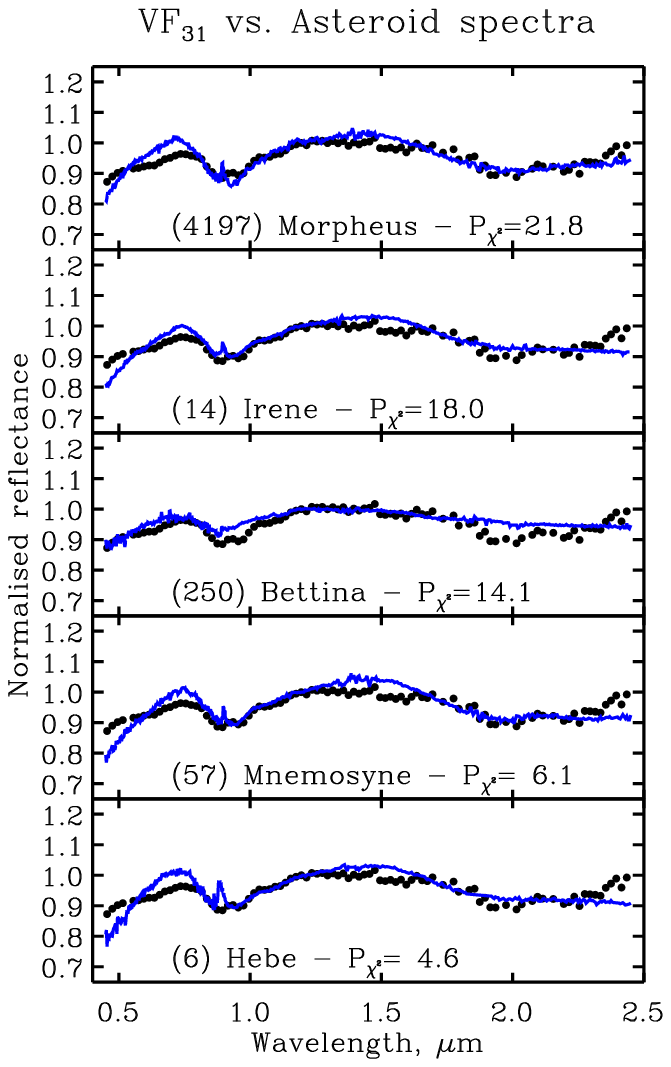}\includegraphics[width=0.33\textwidth,]{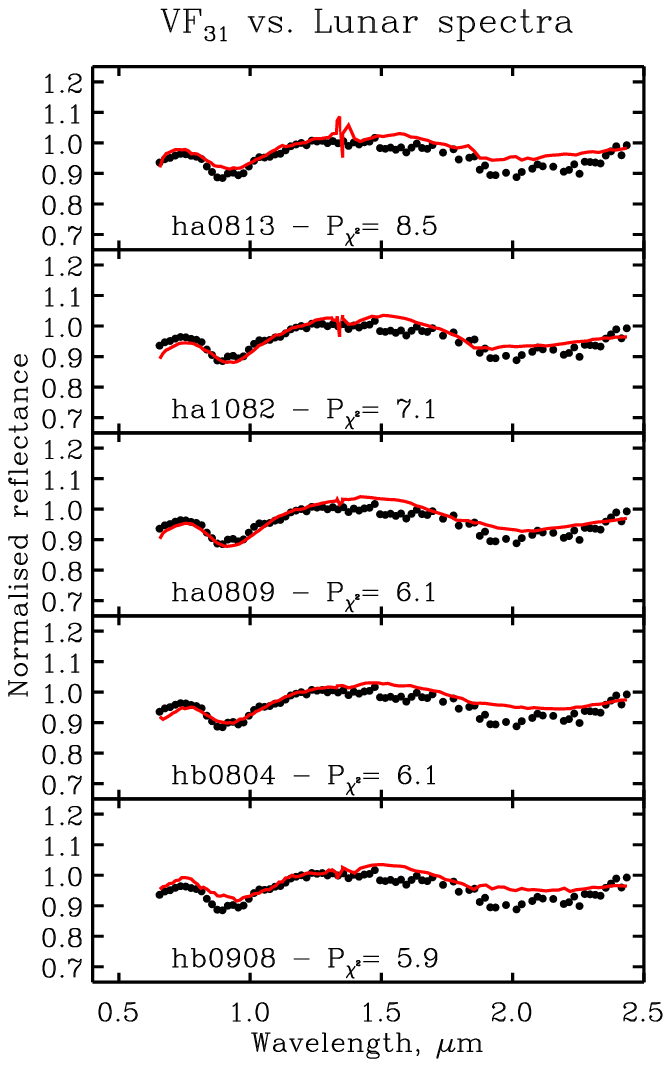}\includegraphics[width=0.33\textwidth]{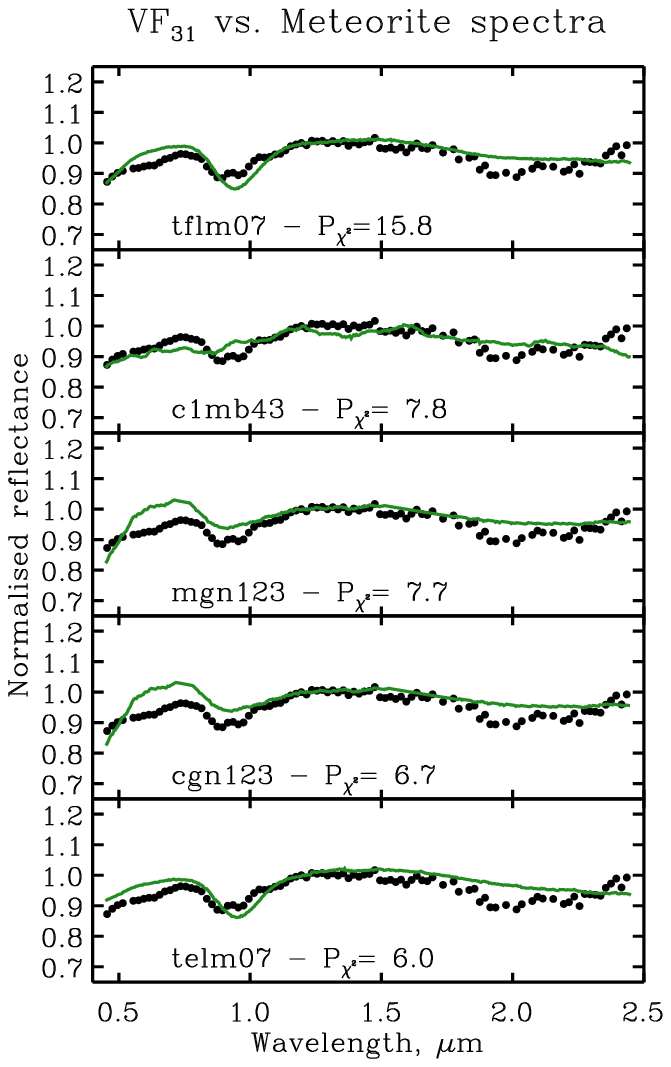}
\caption{Comparison of the VF31 spectrum (black dots) with spectra of different types of objects. 
Left: Asteroid spectra from SMASS II and MITHNEOS. Middle: Lunar surface spectra from PDS. Right: Meteorite spectra from RELAB.}
\label{rankF}
\end{figure*}

Here we utilise asteroid spectra from the Small Main-Belt Asteroid Spectroscopic Survey, Phase II (SMASS II) and the MIT-UH-IRTF Joint Campaign for NEO Spectral Reconnaissance (MITHNEOS)\footnote{\scriptsize http://smass.mit.edu/catalog.php}. 
We considered all asteroid spectra with both visible and near-infrared spectral coverage (427 in total) and applied the same $P$-ranking approach and cut-off criteria as in the previous Section. Top matches are listed in Table~\ref{taxrankast-SR} and shown in the left panel of Figure~\ref{rankF} for the $\chi^2_\mu$ statistic. The reported geometric albedo values are either from IRAS \citep{IRAS} or NEOWISE \citep{NEOWISE}. In addition, we have calculated principal components for these spectra and show them in Fig~\ref{PC1PC2}.

Individual asteroids that spectrally resemble VF31 belong to either the S- or X-complex, although none fit particularly well over the entire spectral region. Interestingly, $P$-ranking for the two scoring metrics highlights the same set of asteroids bar one, suggesting that the scoring is robust against the choice of statistical function.

In fact, the two scoring metrics identify the same top three asteroids: (4197) Morpheus (Sq), (14) Irene (S) and (250) Bettina (Xk). Of those, Bettina lacks the 2\,$\mu$m absorption while Morpheus fits the near-IR part rather well but shows a more pronounced 0.7\,$\mu$m peak than VF31. S-complex members are a better match to VF31's geometric albedo \citep[0.32$^{+0.18}_{-0.11}$;][]{Trilling2007} although the listed X-complex asteroids are within 2-$\sigma$ of the measurement. Based on these considerations, we conclude that our best matches to VF31 are with S or Sq asteroids. 

\begin{figure*}[pos=htb]
\centering
\includegraphics[width=0.7\columnwidth]{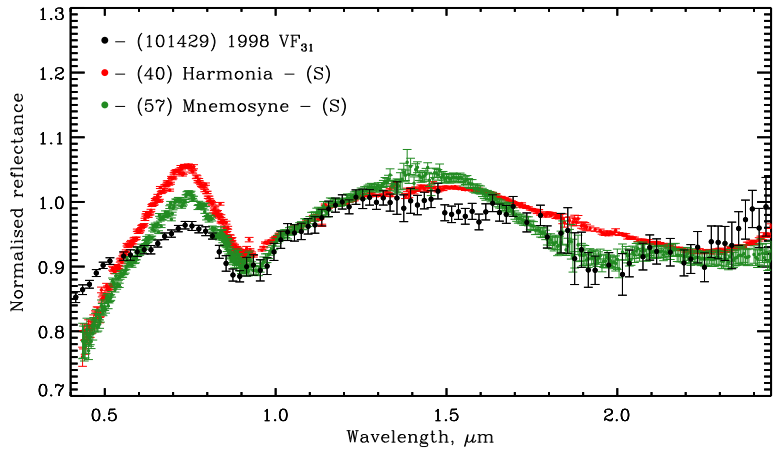}
\caption{Comparison of our VF31 spectrum with the two asteroid analogues from \citet{Rivkin2007}.}
\label{AstRivkin}
\end{figure*}

\citet{Rivkin2007} found that VF31 has a spectrum typical of pyroxene-dominated S-type asteroids classified as S(VII) in the \citet{Gaffey1993} sub-classification scheme. They compared its spectrum with those of two members of this subclass -- (57) Mnemosyne and (40) Harmonia, both S-type asteroids. Mnemosyne is ranked 4th in our $P_{\chi^2}$-sorted list and 5th in our $P_{\Phi_{\rm std}}$-sorted list. We compare the Mnemosyne and Harmonia spectra (green and red points resp.) with that of VF31 in Figure~\ref{AstRivkin}. Mnemosyne is overall the better match but does not fit well in the visible region.

\begin{table*}
\begin{minipage}{\textwidth}
\centering
\caption[Comparison of the VF31 spectrum with lunar spectra.]{Comparison of the VF31 spectrum with lunar spectra.}
\label{LunarSp-SR}
\begin{tabular}{cclcr|cclcr}
\hline 
\hline   
 & & &  Data & $P_{\chi^2_{\mu}}$                                                     &  & & &  Data & $P_{\Phi_{\rm std}}$\\
 ID & Feature & Type${}^{a}$ &  Quality${}^{b}$ & (\%)&  ID & Feature & Type${}^{a}$ &  Quality${}^{b}$ & (\%)  \\
 \hline                 
HA0813 & HADLEY       & MT       & 2 &  8.5 & HA0813 & HADLEY       & MT      & 2  & 16.9\\
HA1082 & LITTROW      & MT       & 1 &  7.1 & HB0908 & CENSORINUS   & CR(H)   & 3  & 10.6\\
HA0809 & ARATUS       & CR(H)    & 1 &  6.1 & HA0819 & APENNINE     & MT      & 1 &  7.7 \\
HB0804 & LANGRENUS    & CF(W)    & 1 &  6.1 & H90426 & DESCARTES    & CR(H)   & 3 &  4.6 \\
HB0908 & CENSORINUS   & CR(H)    & 3 &  5.9 & HB0804 & LANGRENUS    & CF(W)   & 1 &  4.2  \\
HA0819 & APENNINE     & MT       & 1 &  5.7 & HA0847 & BEER         & CR(M)   & 3 &  3.3  \\
HB0812 & DESCARTES    & CR(H)    & 2 &  5.4 & HA1082 & LITTROW      & MT      & 1 &  3.0  \\
HB0926 & PLATO        & CR(H)    & 4 &  4.7 & HC1202 & THEOPHILUS   & CR(H)   & 4 &  3.0  \\
HC1202 & THEOPHILUS   & CR(H)    & 4 &  4.6 & HA0809 & ARATUS       & CR(H)   & 1 &  2.7  \\
HE1186 & FRA MAURO    & CR(H)    & 4 &  3.0 & HB0812 & DESCARTES    & CR(H)   & 2 &  2.4 \\
\hline \noalign{\smallskip}
 \multicolumn{10}{l}{\parbox{140mm}{${}^{a}${CP: Central Peak, CR: Crater, CF: Crater Feature, H: Highlands, W: Wall, MT: Mountain, M: Mare}; ${}^{b}${Ranges from 1=Excellent to 5=Poor}}}
\end{tabular}
\end{minipage}
\end{table*}

\begin{table*}
\centering
\caption[Comparison of VF31 spectrum with RELAB meteorite spectra.]
{Comparison of VF31 spectrum with RELAB meteorite spectra.}
\label{metrank-SR}
\begin{tabular}{l @{\hspace{1.25\tabcolsep}} l @{\hspace{1.25\tabcolsep}} l @{\hspace{1.25\tabcolsep}} l @{\hspace{1.25\tabcolsep}} r|l @{\hspace{1.25\tabcolsep}} l @{\hspace{1.25\tabcolsep}} l @{\hspace{1.25\tabcolsep}} l @{\hspace{1.25\tabcolsep}} r}
\hline 
\hline
  &  Sample  &  &  &  $P_{\chi^2_{\mu}}$                &   &  Sample  &  &  &  $P_{\Phi_{\rm std}}$  \\
ID 		  &  Name		  &  Source		 & Type        & (\%) 	& ID 		  &  Name		  &  Source		 & Type 		& (\%)	\\
\hline
tflm07  &  EETA79001,73    &  Mars   &  Shergottite    & 15.8  & mgn153  & Paragould        &  Other  &  LL5 Chondrite   & 12.1  \\
c1mb43  &  Esquel          &  Other  &  Pallasite      &  7.8  & mgn151  & Paragould        &  Other  &  LL5 Chondrite   & 11.3  \\
mgn123  & Chainpur         &  Other  &  LL3 Chondrite  &  7.7  & cgn151  & Paragould        &  Other  &  LL5 Chondrite   & 11.3  \\
cgn123  & Chainpur         &  Other  &  LL3 Chondrite  &  6.7  & mgn123  & Chainpur         &  Other  &  LL3 Chondrite   & 10.8  \\
telm07  & EETA79001,73     &  Mars   &  Shergottite    &  6.0 & cgn123  & Chainpur         &  Other  &  LL3 Chondrite   &  6.4 \\
mgp098  &  Murray          &  Other  &  CM2	Chondrite  &  5.5 & cgn153  & Paragould        &  Other  &  LL5 Chondrite   &  5.8  \\
mgp086  &  Cold Bokkeveld  &  Other  &  CM2	Chondrite  &  5.4 & tflm07  &  EETA79001,73    &  Mars   &  Shergottite        &  5.1  \\
mgn151  & Paragould        &  Other  &  LL5 Chondrite  &  5.3  & cgp086  &  Cold Bokkeveld  &  Other  &  CM2 Chondrite               &  4.5  \\
cgp098  &  Murray          &  Other  &  CM2 Chondrite  &  4.7  & mgp086  &  Cold Bokkeveld  &  Other  &  CM2 Chondrite               &  4.4  \\
cgn151  & Paragould        &  Other  &  LL5 Chondrite  &  4.3  & cgp098  &  Murray          &  Other  &  CM2 Chondrite               &  3.7  \\
cbmb87  & Y-74659          &  Other  &  Ureilite       &  3.6 & mgp098  &  Murray          &  Other  &  CM2 Chondrite               &  3.7  \\
mgn153  & Paragould        &  Other  &  LL5 Chondrite  & 3.4  & c1tb6   & En95+CaS5        &  Other   & Enstatite &  3.5 \\
cgp086  &  Cold Bokkeveld  &  Other  &  CM2	Chondrite  &  3.4  & sblm03  &  Y-791197,90     &  Moon   &  --                 &  2.7  \\
cgn161  & Rose City        &  Other  &  H5 Chondrite   &  3.2  & telm07  & EETA79001,73     &  Mars   &  Shergottite        &  2.4  \\
mgn161  & Rose City        &  Other  &  H5 Chondrite   &  3.1  & cblm03  &  Y-791197,90     &  Moon   &  --                 &  2.3  \\
c1tb61  & En95+CaS5        &  Other   & Enstatite      &  3.1  & mgn161  & Rose City        &  Other  &  H5 Chondrite                &  2.1  \\
\hline 
\end{tabular}
\end{table*}

\subsubsection{The lunar surface}
\label{seclun}
In our search for good spectral matches we also employed the Tool for Asteroid Modelling -- M4AST\footnote{\scriptsize http://spectre.imcce.fr/m4ast/index.php/index/home} to compare our combined spectrum with laboratory-measured spectra of various materials, lunar samples, meteorites and synthetic laboratory spectra in RELAB.
In the process we found the best matches to be reflectance spectra of Apollo lunar samples, motivating us to investigate this similarity in more detail. One problem in comparing VF31 with lunar samples is that those samples were disturbed when collected from the Moon and brought back to Earth. Therefore we used instead a set of ground-based reflectance spectra of the lunar surface available through the PDS Geoscience Node\footnote{\scriptsize http://pds-geosciences.wustl.edu/missions/lunarspec/}. The data consists of 359 spectra of small lunar areas (3--10\,km across) covering the wavelength range 0.62-2.6\,$\mu$m obtained at Mauna Kea Observatory (MKO) in Hawaii 
\citep{LunarDB1,LunarDB2}. The top matches are listed in Table~\ref{LunarSp-SR} and shown in Figure~\ref{rankF}, middle panel. We observe the same high degree of commonality (8/10 entries) for our two scoring metrics as for the asteroid spectra.
The visual similarity of the VF31 spectrum with the lunar surface spectra is remarkable. We obtain a good fit in the visible part for the feature around 0.75\,$\mu$m, the 1\,$\mu$m and 2\,$\mu$m absorptions and the upturn in the flux beyond 2.1\,$\mu$m. However, the 2\,$\mu$m absorption is somewhat shallower in the lunar data. 

\subsubsection{Meteorites}
\label{secmet}
Finally, we compared the VF31 spectrum with all available meteorite spectra in RELAB. The top hits are listed in Table~\ref{metrank-SR} and presented in the right panel of Figure~\ref{rankF}, where we used the same ranking procedure and cutoff criteria as for the asteroid and lunar spectra. There is again a high degree of commonality (75\% or 12/16 entries) between the two scoring metrics.
Overall, the 2\,$\mu$m feature is either weakly present or altogether missing in the spectra and there is mismatch in the 1\,$\mu$m absorption as well. We also note the presence of several Martian meteorites and of one lunar meteorite in the lists.

Our findings in Sections \ref{asteroids} - \ref{secmet}, taken together, suggest that lunar surface spectra are better matches to the VF31 spectrum than either asteroid or meteorite spectra. We test this further in the next Section.

\begin{table*}
\begin{minipage}{\textwidth}
\centering
\caption[Top-ranked members of the combined set of asteroid, meteorite and lunar surface spectra.]
{Top-ranked members of the combined set of asteroid, meteorite and lunar surface spectra.}
\label{genrank-SR}
\begin{tabular}{lclr|lclr}
\hline 
\hline 
\noalign{\smallskip}
Name & Type${}^1$ & \shortstack[c]{Class} & \shortstack[c]{$P_{\chi^2_{\mu}}$\\ (\%)} &  Name & Type${}^1$ &  \shortstack[c]{Class} & \shortstack[c]{$P_{\Phi_{\rm std}}$\\ (\%)}  \\
\hline  
HADLEY        & LS  & MT    &  7.4  & HADLEY       & LS  & MT     &  15.3  \\
LITTROW       & LS  & MT    &  5.9  & CENSORINUS      & LS  & CR(H)  &   9.3  \\
LANGRENUS   & LS  & CF(W) &  5.2  & APENNINE  & LS  & MT     &   6.8  \\
CENSORINUS       & LS  & CR(H) &  5.1  & DESCARTES     & LS  & CR(H)  &   4.1  \\
ARATUS           & LS  & CR(H) &  5.1  & LANGRENUS  & LS  & CF(W)  &   3.5  \\
APENNINE   & LS  & MT    &  5.0  & BEER            & LS  & CR(M)  &   2.8  \\
DESCARTES     & LS  & CR(H) &  4.3  & (14) Irene      & Ast & S      &   2.7  \\
(4197) Morpheus  & Ast & Sq    &  4.2  & THEOPHILUS & LS  & CR(H)  &   2.5  \\
(14) Irene       & Ast & S     &  4.2  & LITTROW      & LS  & MT     &   2.5  \\
PLATO            & LS  & CR(H) &  3.8  & ARATUS          & LS  & CR(H)  &   2.4  \\
THEOPHILUS  & LS  & CR(H) &  3.8  & (250) Bettina   & Ast & Xk     &   2.2 \\
\hline \noalign{\smallskip}
 \multicolumn{7}{l}{\parbox{87mm}{$^1$ Ast: Asteroid, Met: Meteorite, LS: Lunar surface}}
\end{tabular}
\end{minipage}
\end{table*}

\subsubsection{Ensemble ranking}
Here we compare the VF31 spectrum with all types of available spectra simultaneously: asteroid, lunar surface or meteorite. To maintain consistency, we recompute $P_{\chi^2_{\mu}}$ and $P_{\Phi_{\rm std}}$ values for a common wavelength range (0.656--2.455\,$\mu$m) and resample to the same number of data points (93). We note that use of a common wavelength range will generally change the values of $P$ from those presented in Tables \ref{taxrankast-SR}, \ref{LunarSp-SR} and \ref{metrank-SR}. 
In Table~\ref{genrank-SR} we present $P$-ranking scores for the combined sample, showing all hits with $P_{\chi^2_{\mu}}>$3\,\% and the same number of elements for $P_{\Phi_{\rm std}}$. The lists are dominated by lunar spectra (8 out of 10 listed matches) and complemented by some of the asteroid entries in Table~\ref{taxrankast-SR}, appearing in the bottom half of either list. All meteorite spectra score less than $3$\% for either statistic, therefore meteorite matches do not appear in these lists. The outcome of this exercise reinforces our earlier conclusion that lunar surface data are the best spectral match to VF31.

The top-ranked lunar surface spectra are from crater interiors and mountain features in the lunar highlands. Their composition is mainly anorthite (CaAl$_2$Si$_2$O$_8$) which is rare on the Earth as well as different kinds of magnesoferrous (``mafic'') minerals, mainly pyroxene and olivine. Depending on the abundance of these minerals the 2\,$\mu$m absorption feature can be suppressed by different amounts \citep{AsteroidsIII}. The weakness of the absorptions in the VF31 spectrum, with respect to other asteroids, may be due to the presence of agglutinates, which degrade diagnostic spectral absorptions, or the result of space weathering. The type of pyroxene present is probably responsible for the position of the 1\,$\mu$m feature. Low-Ca pyroxene (e.g.~orthopyroxene - OPX) has an absorption at 0.93--0.95\,$\mu$m and high-Ca pyroxene (e.g.~clinopyroxene - CPX) at 0.95-1.00\,$\mu$m \citep{Yong-Liao2004}. The presence of olivine broadens the former feature and suppresses the 2\,$\mu$m feature as well. The position of the 1\,$\mu$m feature observed in the spectrum of VF31 is consistent with OPX while the presence of the 2\,$\mu$m absorption indicates a moderate pyroxene abundance. 
\begin{figure}[pos=h]
\centering
\includegraphics{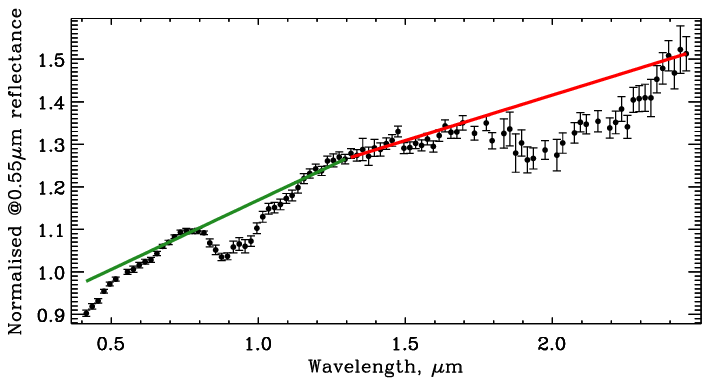}\\
\includegraphics{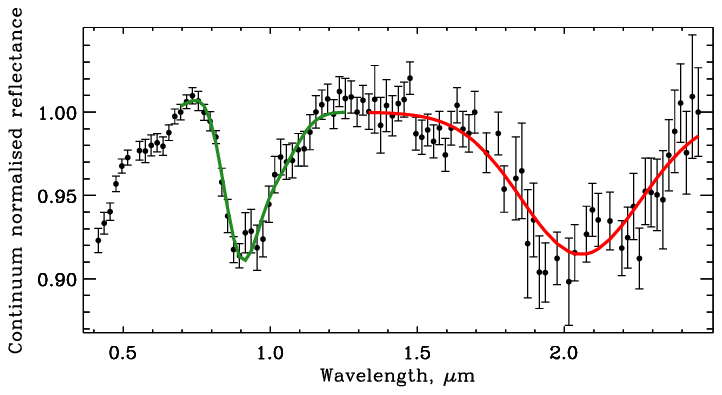} 
\caption{{\it Top:} Continuum fitting at the 1 $\mu$m (green line) and 2 $\mu$m (red line) absorptions. {\it Bottom:} Fits to the continuum-normalised spectrum around the 1 and 2 $\mu$m features with Gauss-Hermite orthogonal functions (see text for details).}
\label{VF31_1m2m}
\end{figure}
\begin{figure}[pos=h]
\centering
\includegraphics[width=0.4\columnwidth]{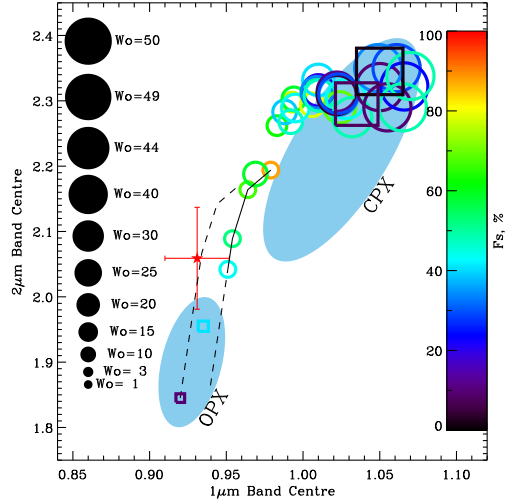} 
\caption{Position of 1 and 2 $\mu$m absorptions for VF31 (red star) and formal 3-$\sigma$ uncertainties compared with RELAB data for pyroxene with different Wo and Fs numbers from \citet{OPX} and \citet{Horgan2014}, represented by circles and squares respectively.}
\label{OPX_CPX}
\end{figure}

To obtain the positions of the absorption minima in our spectrum we follow the approach by \citet{gauss-hermite} for identification of non-gaussian line profiles, using Gauss-Hermite series of orthogonal functions to fit the spectral lines. These are normalised gaussian functions multiplied by the linear combination of the 3$^{\rm rd}$ and 4$^{\rm th}$ order Hermite polynomials given by the following expressions: 
\begin{eqnarray}
{\rm G}(y)& = & a{1 \over\sigma\sqrt{2\pi}}{\rm e}^{-{1 \over 2}y^2} \\
{\rm H_3}(y)&=&h_3{1\over \sqrt{3}}(2y^3-3y) \\
{\rm H_4}(y)&=&h_4{1\over \sqrt{24}}(4y^4-12y^2+3) \\
{\rm G}_{H}&=&{\rm G(1+H_3+H_4)}
\label{GH}
\end{eqnarray}
where $y=\left({x-\mu}\right)/\sigma$. 
The advantage of this approach is that it yields weakly-correlated parameter estimates \citep{GH}. Here we perform a Levenberg-Marquardt nonlinear-least-squares fit \citep{mpfit} separately to the 1 and 2\,$\mu$m features with the $G_{H}$ function and obtain estimates for the set of coefficients \mbox{\{$a, \mu, \sigma, h_3, h_4$\}} and formal 1-$\sigma$ uncertainties. We are interested in the $\mu$ values, which represent the position of minimum absorption for the 1 and 2 \,$\mu$m features. In our case they are 0.931$\pm$0.007 and 2.059$\pm$0.026\,$\mu$m (Figure~\ref{VF31_1m2m}) respectively, which is more consistent with pyroxene low in Ca, e.g.~OPX \citep{Horgan2014}. 

We compare the positions of the 1 and 2\,$\mu$m features with data in \citet{OPX} for which the Ca, Mg, Fe abundance is known. This sample is populated solely by CPX with Ca abundance (Wo number) $\geq$10\% and we complement this with data from \citeauthor{Horgan2014} with low Wo number. The comparison is shown in Figure~\ref{OPX_CPX} where our measurement for VF31 is indicated by a red star. Data are represented as circles with different sizes to indicate Ca content and different colour to indicate Fe content. For measurements with low Ca content, i.e. Wo=10, there is sufficient data to indicate a trend with changing Fe abundance (solid line). If we extrapolate this trend to even lower Ca content (Wo=1; dashed line) we find that VF31 lies near this trend line, implying that its surface is highly abundant in OPX with moderate Fe abundance. Even if the uncertainty in the locations of the absorptions is underestimated by a factor of 3, VF31 is more consistent with OPX than CPX, especially in terms of the 1\,$\mu$m feature (Figure~\ref{OPX_CPX}).

There are several factors affecting the spectral profile of an asteroid surface: space weathering \citep{AsteroidsIII}; regolith particle size, where smaller particles redden the slope and decrease the depth of absorption features \citep{ParSize}; a high content of iron or other mineral such as plagioclase; and phase reddening \cite[PR; e.g.][]{PhaseRed}.
PR is relevant to VF31 because of the differing phase angles of the VIS and NIR observations for our composite spectrum (26$^\circ$ and 42$^\circ$ resp.; Table~\ref{phaseang}). However, PR strongly depends on taxonomic type, with olivine-rich asteroids being particularly susceptible to it \citep{Perna2018}. Because we do not see evidence of PR on spectra of the olivine-dominated Eureka family observed at 2$^\circ$-27$^\circ$ phase angle (see Section~\ref{sec:eurekafamily}), we believe our spectrum of VF31 is not significantly affected either.

With the information gathered so far, we can perform mixture fitting to the VF31 spectrum. In particular, we use the spectral slope to constrain the combination of space weathering, iron content and/or particle size that best matches the spectrum. Our analysis so far strongly indicated the presence of OPX; we adopt this mineral as one of the baseline endmembers in this exercise.
\begin{table*}
\centering
\caption[Results from mixture fitting of the VF31 spectrum. Entries in bold are also shown in Fig.~\ref{MixFits4}.]{Results from mixture fitting of the VF31 spectrum. Entries in bold are also shown in Fig.~\ref{MixFits4}.}
\label{MixFits}
\begin{tabular}{lrrrr}
\hline 
\hline      
Mineral  &  Proportion  &  SW                    &  $\chi^2$  &  Albedo \\                     or Mixture         & \%                & wt\% npFe   &                  &              \\
\hline
\multicolumn{5}{c}{\it Single minerals} \\
\hline
{\bf OPX080}  &  {\bf 100}  &  {\bf 0.040}  &  {\bf 23.6614}  &  {\bf 0.20} \\
{\bf OPX250}  &  {\bf 100}  &  {\bf 0.140}  &  {\bf 26.0996}  &  {\bf 0.14} \\
{\bf OPX975}  &  {\bf 100}  &  {\bf 0.010}  &  {\bf 5.3495}  &  {\bf 0.50} \\
OLVFo11  &  100  &  0.022  &  19.6011  &  0.23 \\
OLVFo90  &  100  &  0.017  &  6.3926  &  0.48 \\
Odessa Iron Meteorite  &  100  &  0 &   2.9552  &  0.13 \\
Plagioclase (PGC)  &  100  &  0.009  &  2.7731  &  0.64 \\ \hline
\multicolumn{5}{c}{\it Two-mineral mixtures} \\ \hline
{\bf OPX080+Odessa}  &  {\bf 07/93}  &  {\bf 0}  &  {\bf 2.1076}  &  {\bf 0.13} \\
OPX080+PGC  &  02/98  &  0.009  &  1.8963  &  0.59 \\
{\bf OPX250+Odessa}  &  {\bf 05/95}  &  {\bf 0.005}  &  {\bf 2.1378}  &  {\bf 0.13} \\
OPX250+PGC  &  01/99  &  0.011  &  1.8643  &  0.58 \\
{\bf OPX975+Odessa}  &  {\bf 61/39}  & {\bf  0.015}  &  {\bf 1.2762}  & {\bf 0.24} \\
OPX975+PGC  &  19/81  &  0.009  &  1.4207  &  0.59 \\  \hline
\multicolumn{5}{c}{\it Three-mineral mixtures} \\ \hline
{\bf OPX080+OLVFo90+Odessa}  &  {\bf 05/24/71}  &  {\bf 0.014}  &  {\bf 1.9456}  &  {\bf 0.17} \\
{\bf OPX080+PGC+Odessa} &  {\bf 06/56/38}  &  {\bf 0.018}  &  {\bf 1.2710}  &  {\bf 0.25} \\
OPX170+OLVFo90+Odessa   & 05/27/68 & 0.018 & 1.9309  & 0.17 \\
OPX170+PGC+Odessa       & 05/60/35 & 0.021 & 1.1878  & 0.25 \\
{\bf OPX250+OLVFo90+Odessa}   & {\bf 04/27/69} & {\bf 0.019} & {\bf 1.9102}  & {\bf 0.17} \\
{\bf OPX250+PGC+Odessa}       & {\bf 04/58/38} & {\bf 0.020} & {\bf 1.2069}  & {\bf 0.24} \\
OPX500+OLVFo90+Odessa   & 11/27/62 & 0.021 & 1.7742 & 0.18 \\
OPX500+PGC+Odessa       & 11/59/30 & 0.020 & 1.0407 & 0.27 \\
{\bf OPX975+OLVFo90+Odessa}  &  {\bf 57/07/36}  & {\bf 0.016}  &  {\bf 1.2532}  &  {\bf 0.25} \\
{\bf OPX975+PGC+Odessa} &  {\bf 46/30/24}  &  {\bf 0.016}  &  {\bf 1.1432}  &  {\bf 0.31} \\ \hline
\multicolumn{5}{c}{\it Four-mineral mixtures} \\ \hline
\noalign{\smallskip}
{\bf OPX080+OLVFo90+PGC+Odessa} &   {\bf 06/00/56/38}  &   {\bf 0.018}  &  {\bf 1.2727}  &   {\bf 0.25} \\
{\bf OPX250+OLVFo90+PGC+Odessa} &   {\bf 04/00/57/39}  &   {\bf 0.023}  &  {\bf 1.2125}  &   {\bf 0.24} \\
 {\bf OPX975+OLVFo90+PGC+Odessa} &   {\bf 44/04/30/22}  &   {\bf 0.016}  &  {\bf 1.1315}  &   {\bf 0.32} \\
\hline
\end{tabular}
\end{table*}

\begin{figure*}[pos=h!]
\includegraphics[width=0.32\textwidth,angle=0]{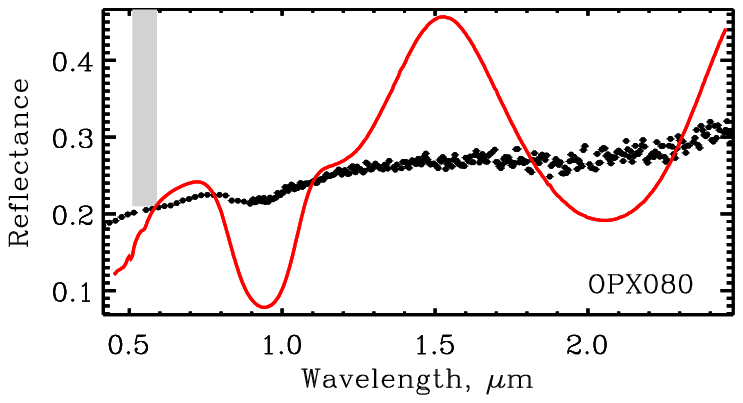}
\includegraphics[width=0.32\textwidth,angle=0]{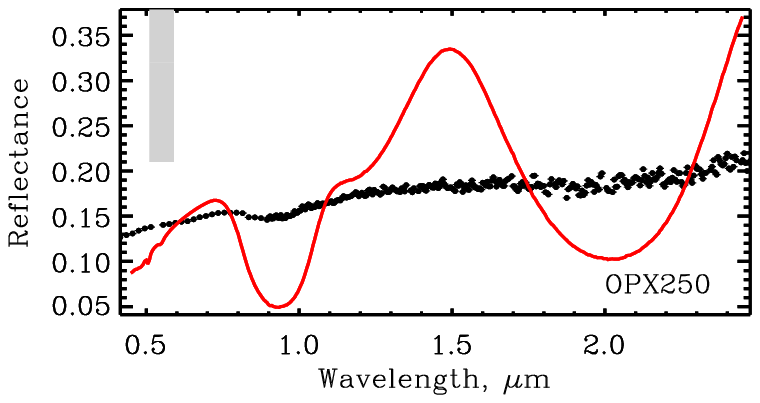}
\includegraphics[width=0.32\textwidth,angle=0]{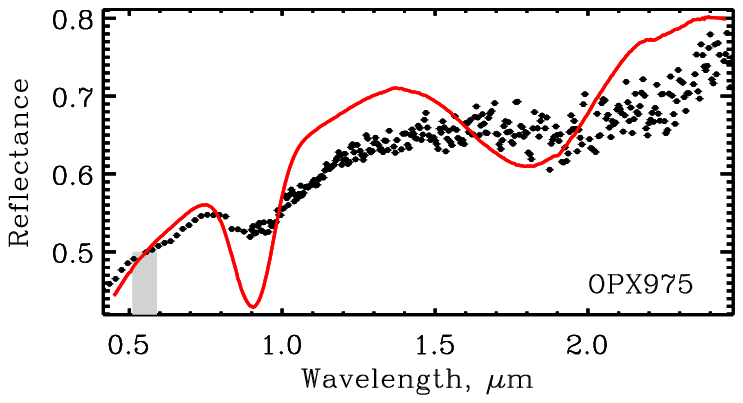}\\
\includegraphics[width=0.32\textwidth,angle=0]{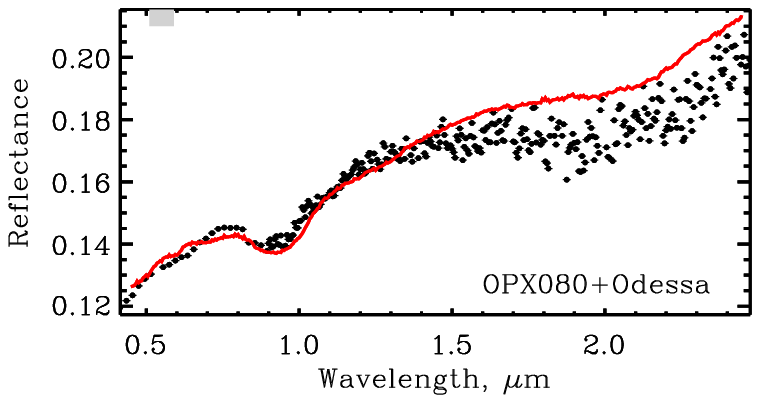}
\includegraphics[width=0.32\textwidth,angle=0]{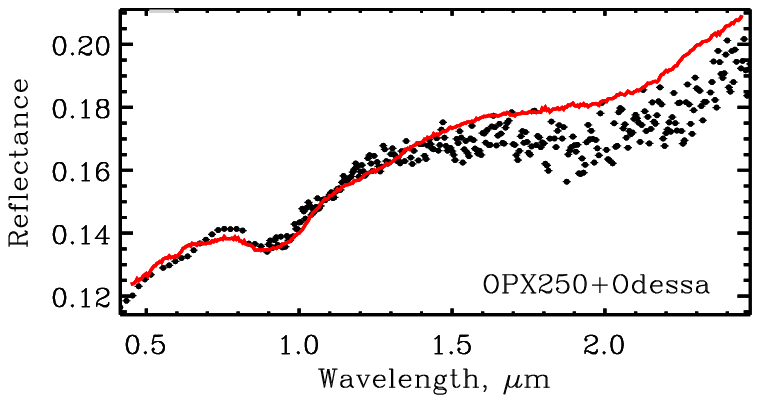}
\includegraphics[width=0.32\textwidth,angle=0]{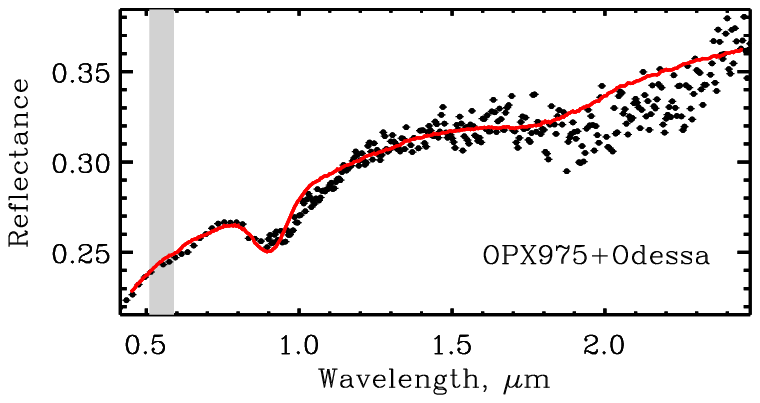}\\
\includegraphics[width=0.32\textwidth,angle=0]{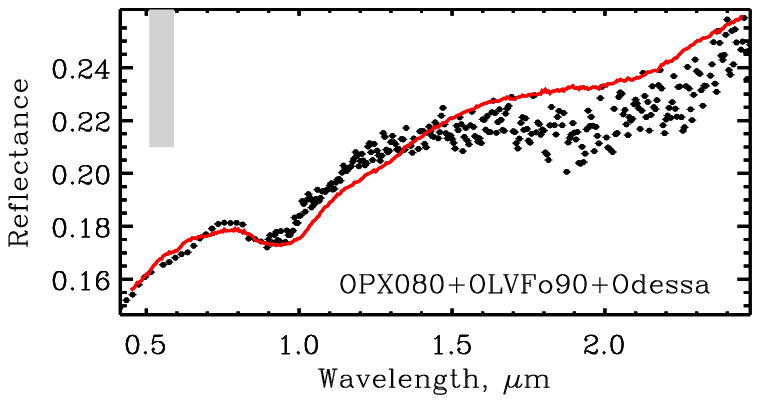}
\includegraphics[width=0.32\textwidth,angle=0]{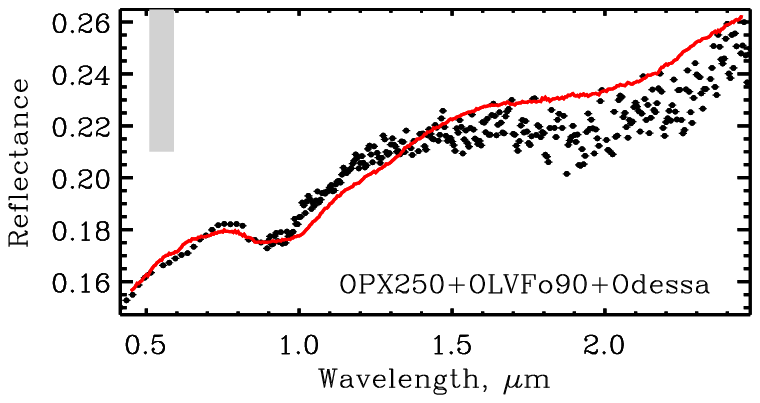}
\includegraphics[width=0.32\textwidth,angle=0]{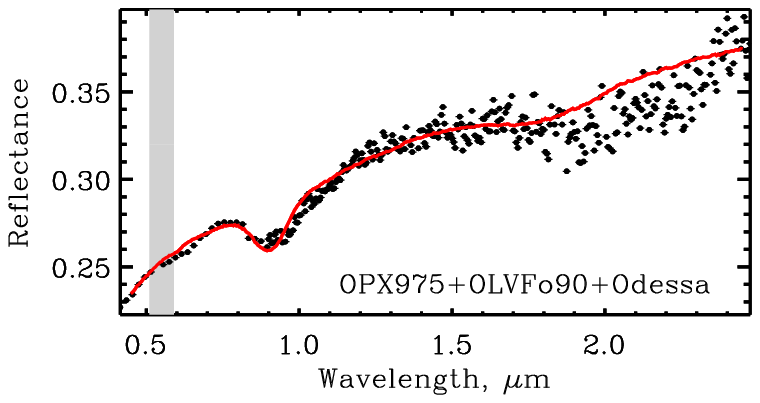}\\
\includegraphics[width=0.32\textwidth,angle=0]{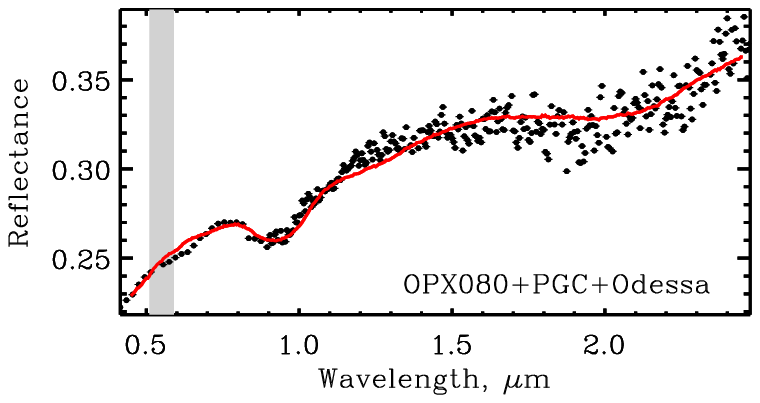}
\includegraphics[width=0.32\textwidth,angle=0]{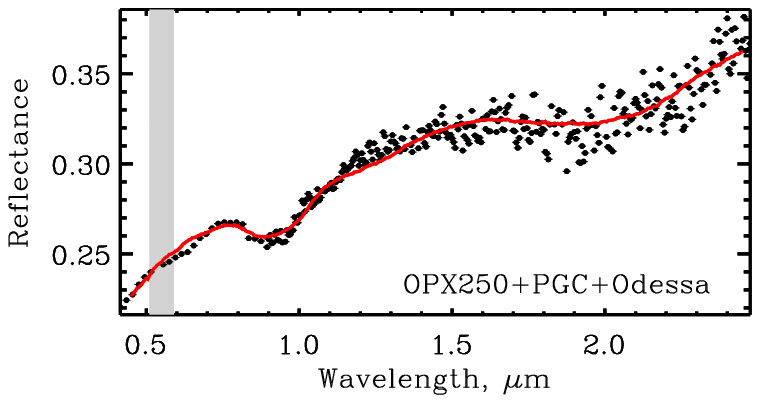}
\includegraphics[width=0.32\textwidth,angle=0]{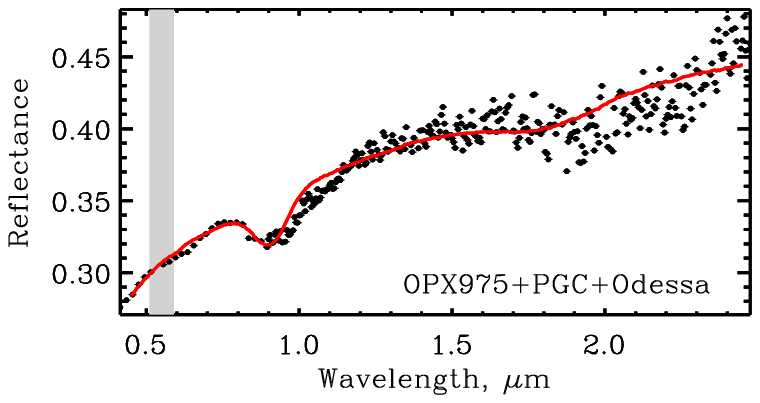}\\
\includegraphics[width=0.32\textwidth,angle=0]{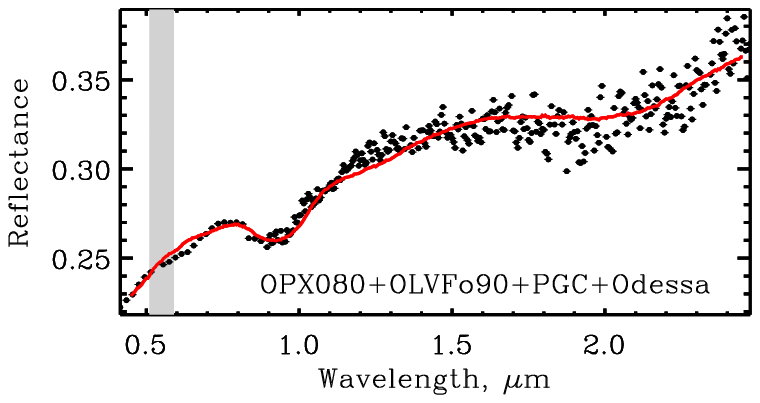}
\includegraphics[width=0.32\textwidth,angle=0]{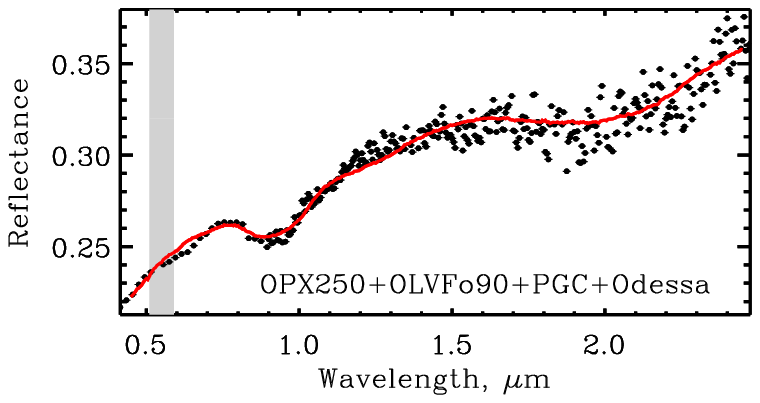}
\includegraphics[width=0.32\textwidth,angle=0]{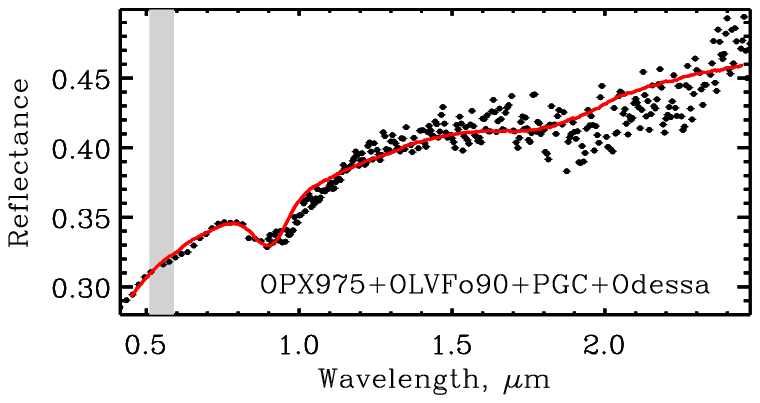}
\caption{Selected mixture models for VF31 from Table~\ref{MixFits}. The number of elements increases from top to bottom. The grey vertical band indicates the 1-$\sigma$ range in the albedo.}
\label{MixFits4}
\end{figure*}

\subsubsection{Mixture fitting}
We model the spectrum of VF31 using candidate mineral components in a simplified Hapke spectral mixing model \citep{Hapke1981,Hapke1984,Hapke1986}. In the case of intimately mixed material, such as asteroid regolith, the observed spectrum does not correspond to the linear combination of the individual component spectra. However, using the Hapke reflectance theory one can convert the reflectance spectrum ($r$) of individual components to their single scattering albedo ($w$) which, under Hapke theory, can be linearly combined. The linear combination of single scattering albedo can then be transformed back to the reflectance spectrum and compared to the asteroid spectrum. 

We combine spectra of several candidate end-members to find satisfactory fits to the asteroid spectrum. From the fitting procedure, we obtain the relative abundance of each end-member. Unlike the $P$-ranking described in previous Sections, here we use the unbinned spectral data for VF31 in the fits. In addition, we account for space weathering using the \citet{Hapke2001} procedure by adding npFe nanophase iron within the host material while taking into account the optical constants of iron. Details on the mixing and space-weathering procedures applied to asteroids can be found in \citet{Max2018}.

Selected fits from our mineralogical analysis are listed in Table~\ref{MixFits} with items in bold also shown in Fig~\ref{MixFits4}. First, we compared the observed spectrum of VF31 with individual end-member spectra from RELAB. These included: orthopyroxene (OPX), olivine (OLV), plagioclase (PGC) and the iron meteorite MET101A (Odessa). We find that Mg-rich mineral endmembers (OPX97.5, OLVFo90) are better matches than iron-rich endmembers (OPX08, OLVFo11) though the fits are generally unsatisfactory and with too high albedos. The best single-endmember fits were achieved by either iron meteorite or plagioclase, however the high albedo of the latter  and low albedo of the former are inconsistent with the VF31 measurement. 
\begin{figure*}[pos=h]
\centering
\includegraphics[width=0.6\columnwidth]{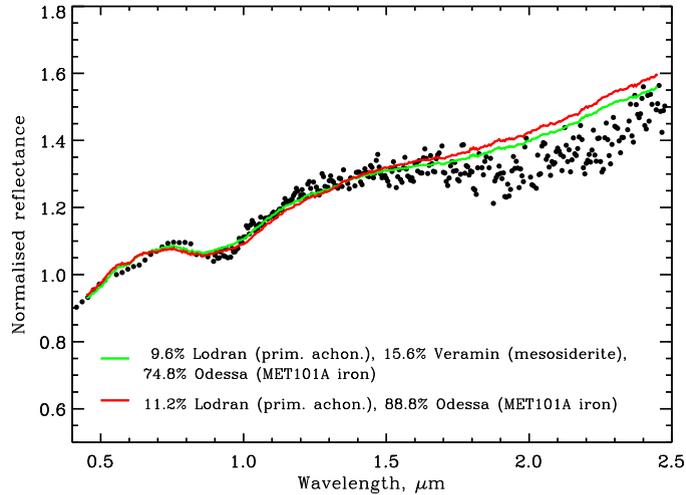} 
\caption{Mixture fitting with achondrite meteorite endmembers from \citet{Rivkin2007} and including the Odessa iron meteorite.}
\label{Rivkin_1_2met}
\end{figure*}

Next, we consider mixtures of two or more elements. Here it is important to note that, for groups with a fixed number of elements, the $\Delta \chi^2$ value at even a 1-$\sigma$ level of significance is $>$2. Therefore, the information on Table~\ref{MixFits} should be viewed as groups of potential matches, similar to the $P$-ranking approach used earlier in the paper, where all mixtures in the same group are statistically equivalent. The best-fit mixtures among the different groups are determined by visual inspection, taking into account the asteroid albedo and the number of elements in the mixture.

 Our best-fit two-element mixture is OPX with high (97.5\%) Mg content combined with iron meteorite in proportions of 61\% and 39\% respectively (right panel, second from top in Figure~\ref{MixFits4}). This mixture is also consistent with the albedo at the 1-$\sigma$ level. As a simple robustness test of the fit, we replaced the Odessa spectrum with the average for all available iron meteorite spectra in RELAB. This does not noticeably alter the result and we retained use of the Odessa spectrum for the remainder of the exercise. 
 Reducing the Mg content in OPX results in the fits shown in the left and middle panels. These are better matches to the shape and location of the 2\,$\mu$m absorption though they also show a higher slope in that region. However, they are inconsistent with the asteroid albedo and we find this to be true also if we replace the iron meteorite with other endmembers. 
 
\begin{figure*}[pos=h]
\centering
\includegraphics[width=0.5\columnwidth]{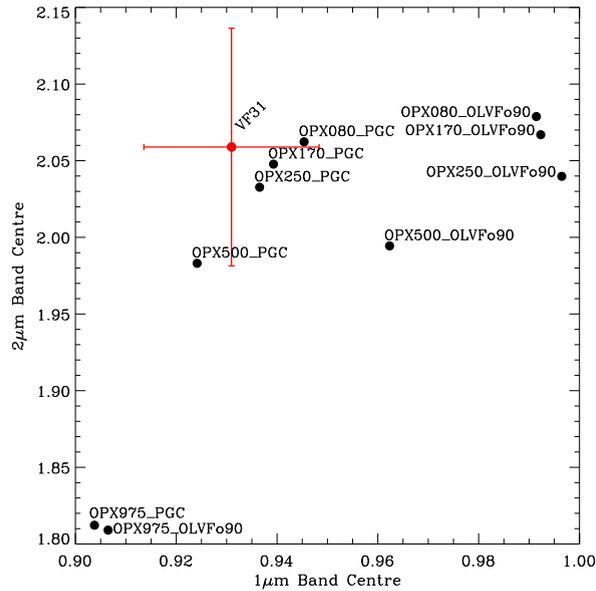} 
\caption{Positions of 1 and 2\,$\mu$m absorptions of OPX three-element mixtures containing olivine, plagioclase, and meteoritic iron compared to VF31.}
\label{OPXs_1m2m}
\end{figure*}

If we allow a third element in the mixture, we find that olivine has little effect on the spectral profile but adding plagioglace results in visually satisfactory fits for mixtures with low-to-moderate OPX that are also consistent with the albedo. Interestingly, these mixtures are dominated by the PGC and iron metal components (94-96\%) rather than OPX (4-6\%). In contrast, Mg-rich OPX accounts for 50\% or more in the respective three-element mixture. The presence of plagioclase would actually be consistent with the spectral similarity of VF31 to the lunar surface (Section~\ref{seclun}). Plagioclase was also identified as one of the main surface constituents of asteroid Itokawa through examination of material returned to Earth by the Hayabusa spacecraft \citep{Hayabusa}. The effect of OPX composition on absorption band location for the three-element mixtures can be better illustrated by plotting one band against the other. We have calculated the 1 and 2\,$\mu$m band locations for the different mixtures in the same way as we did for VF31 in the previous Section and plotted them in Figure~\ref{OPXs_1m2m}. The error bars for VF31 are 3$\times$ the formal uncertainty. Clearly, three-element mixtures that include OPX at 8-50\,\% Mg abundance and plagioclase are more compatible with VF31 than mixtures with either olivine or Mg-rich OPX. 

Allowing both PGC and olivine in a four-element mixture and re-optimising (bottom row) has the effect of pushing the olivine out of the mixture and we recover the OPX-PGC-Odessa solution, in other words there is no further improvement in fit quality. All mixtures shown in Figure~\ref{MixFits4} yield similar abundances of npFe, 1.4-2.3\% by weight, with the exception of the two-element low-to-moderate OPX mixtures (0-0.5\% by weight).

For completeness, we have also fit the spectrum to two mixtures with meteorite material presented in \citet{Rivkin2007} -- Lodran (primitive achondrite), Veramin (mesosiderite) and Odessa (iron) (Figure~\ref{Rivkin_1_2met}). In those cases, the spectral profile shortward of the 1\,$\mu$m absorption in the meteorite mixture is shifted to shorter wavelengths and the 2\,$\mu$m feature becomes quite shallow. The fits are somewhat similar to the two-element low-OPX mixtures in Fig.~\ref{MixFits4}.
The $\chi^{2}$ values obtained here are 1.68 and 2.03 for the 3- and 2- member mixtures respectively and with albedos of 0.13 and 0.12 which are rather low for VF31.

In conclusion, the simplest mixture which adequately represents the spectrum of VF31 is Mg-rich orthopyroxene mixed with iron. The most promising alternative appears to be a mixture of iron and plagioclase with a small amount of Mg-poor orthopyroxene. This mixture better reproduces the shape and location of the 2\,$\mu$m absorption at the expense of introducing an extra degree of freedom to the model. Though based on the best information available at this time, our mixture fitting results should be treated with some caution as they depend on combining measurements from different spectral regions obtained under different sets of circumstances. For this reason, we have strived to describe our data reduction and analysis in detail so that its advantages or shortcomings are clear to the reader.

\section{\label{sec:eurekafamily} Eureka family asteroids}
\begin{figure*}[pos=h!]
\centering
\includegraphics[width=0.6\columnwidth]{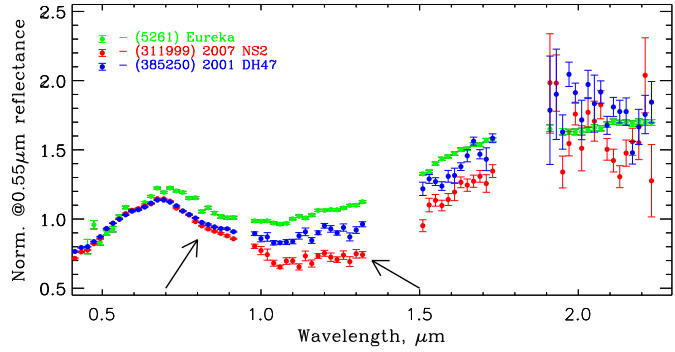} 
\caption{Reflectance spectra of (5261) Eureka, (311999) 2007 NS$_2$ and (385250) 2001 DH$_{47}$. The ``shoulders'' referred to in the text are marked with arrows.}
\label{Sp3MarTr}
\end{figure*}

Eureka family asteroids (311999) 2007 NS$_2$ and (385250) 2001 DH$_{47}$, referred to as NS2 and DH47 hereafter, were observed with X-SHOOTER (Table~\ref{phaseang}). Their spectra, together with a (5261) Eureka spectrum obtained in 2005 as part of the the MIT-Hawaii Near-Earth Object Spectroscopic Survey (MITHNEOS) are presented in Figure~\ref{Sp3MarTr}. All spectra have been normalised to 0.55\,$\mu$m. 

\begin{figure}[pos=h!]
\centering
\includegraphics{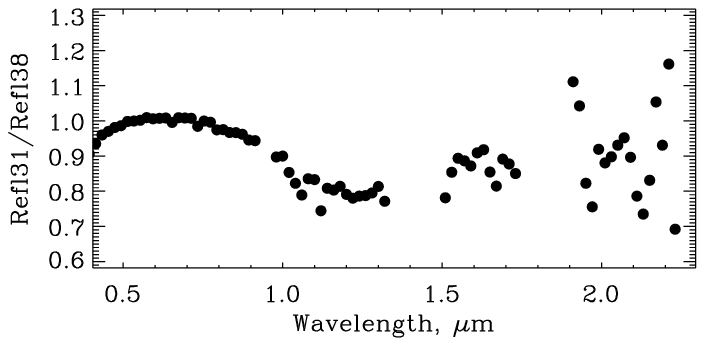}\\
\includegraphics{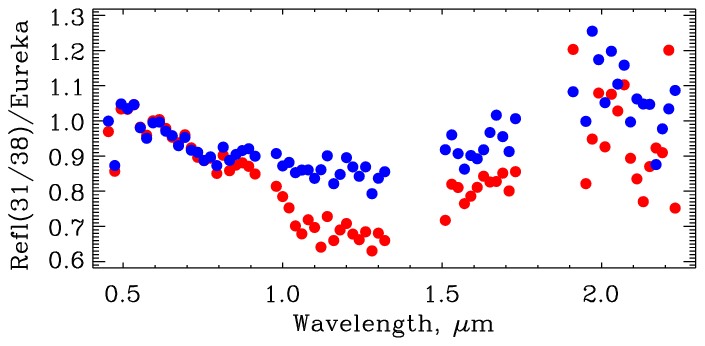} 
\caption{Comparison of spectra of Eureka family members. {\it Top:} NS2 divided by DH47. Note the coincidence in the spectral profiles up to $\sim$0.8$\mu$m. {\it Bottom:} NS2 (red) and DH47 (blue) divided by Eureka.}
\label{SpComp}
\end{figure}

The spectra of NS2 and DH47 virtually coincide in the range 0.52-0.78\,$\mu$m and diverge from $\sim$0.8\,$\mu$m onwards. This can be more clearly seen by plotting the spectral ratio (Figure~\ref{SpComp}, top panel). Eureka's spectrum is indistinguishable, within uncertainties, from the other two up to $0.7$\,$\mu$m (bottom panel) but shows a higher relative reflectance than either asteroid longward of that wavelength. 

There is a drop in the reflectance of NS2 relative to DH47 between 1 and 1.05\,$\mu$m. A similar feature, although less pronounced, may be present in the spectrum of DH47 as well. The slope behaviour of the two spectra is very similar between 1 and 1.7\,$\mu$m. Over the same range the relative reflectance of NS2 is lower than that of DH47.

Although space weathering may be causing some of the differences in the spectra, it is probably not solely responsible for them. Laboratory experiments show that space weathering in olivine (in the form of npFe formed by heating) changes the slope over the entire spectrum, not preferentially the red or blue part (eg \citet{Kohout2014}, their Figure 6, bottom panel; also \citet{Kaluna2017}). Production of npFe by heating (\citeauthor{Kohout2014}), laser irradiation \citep{Yamada1999,Sasaki2001Nature}, irradiation by H$^+$, He$^+$, Ar$^+$, and Ar$_2$ \citep{Brunetto2005} change spectra in different ways but all show reddening (see also Figure~1 in \citet{Sasaki2001LPSC}). Olivine appears to be more easily weathered than pyroxene \citep{Yamada1999,Hiroi2001}. 

If the differences in the two spectra are not solely due to space weathering, they can be due to any combination of space weathering with: (a) phase reddening (b) different particle size distribution, or (c) composition. We discuss each process separately below.
\paragraph{Phase reddening:} 
Phase reddening increases the spectral slope and decreases the strength of the absorption features with increasing phase angle. \citet{PhaseRed} found differences in the spectral slope of NEAs observed at different phase angles. The same authors obtained laboratory spectra of olivine-rich LL6 chondrite material at different phase angles and found significant differences in the near-IR for phase angles $>$30$^\circ$ \cite[cf their Figures 5 and 6; see also][]{Perna2018}. Our observations were obtained at phase angles $<$30$^\circ$ (Table~\ref{phaseang}) therefore our spectra should not be strongly affected by phase reddening. Indeed, we expect the spectral slope should increase with the phase angle, but here the opposite is true, i.e.~the spectral slope for Eureka is \mbox{$0.3329\pm0.0008$} $\mbox{$\mu$m}^{-1}$ (phase angle $\sim$4.5$^\circ$), \mbox{$0.1818\pm0.0032$} $\mbox{$\mu$m}^{-1}$ for DH47 (phase angle $\sim$13$^\circ$) and \mbox{$0.1439\pm0.0037$} $\mbox{$\mu$m}^{-1}$ for NS2 (phase angle $\sim$26$^\circ$).
\begin{figure*}[pos=h]
\centering
\includegraphics[width=0.4\columnwidth]{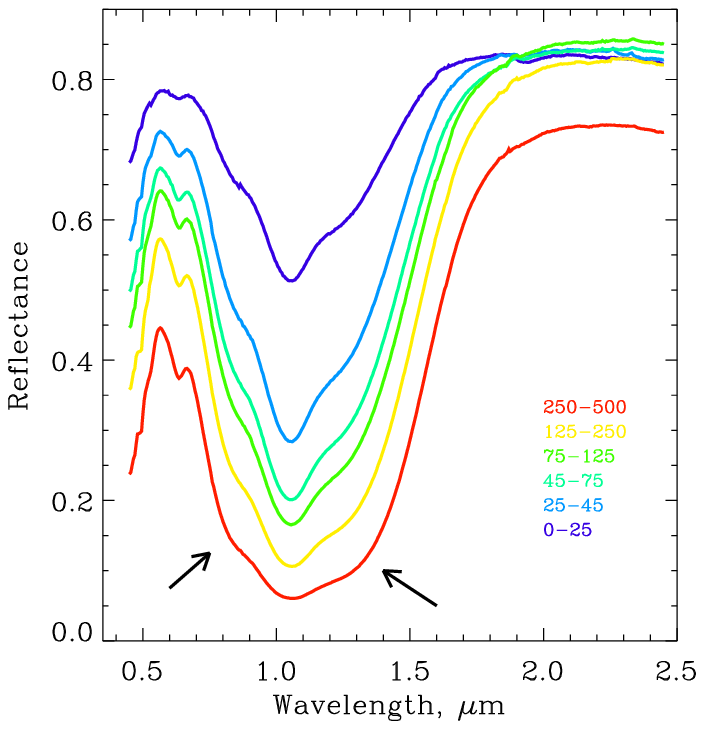}\hspace{-0.5cm}\includegraphics[width=0.4\columnwidth]{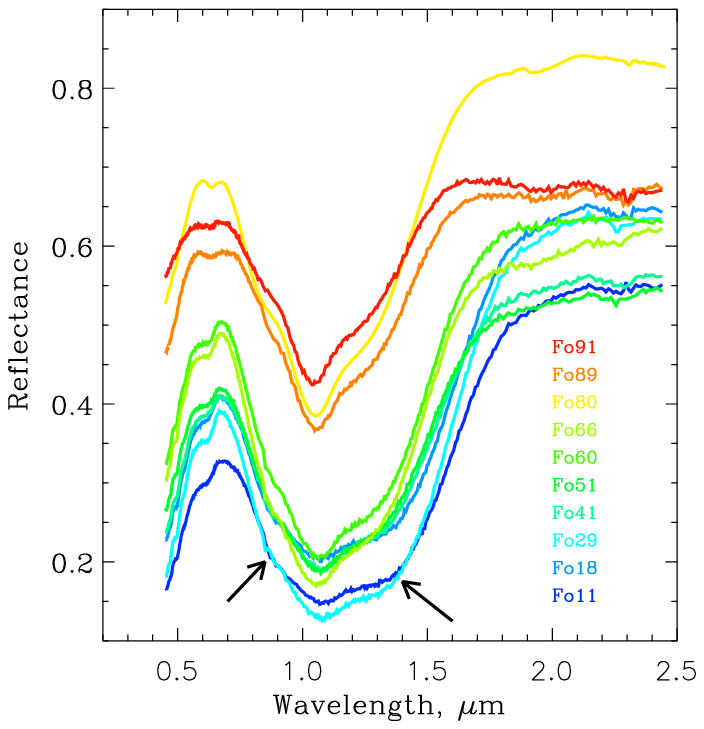} 
\caption{{\it Left:} Spectra of particulate olivine with different particle size. {\it Right:} Spectrum of olivine with different Fo number. The shoulders referred to in the text are marked with arrows.}
\label{FoXX_Size}
\end{figure*}

\paragraph{Surface particle size:} 
\citet{Cloutis2015} show that the 1\,$\mu$m absorption in samples of particulate olivine broadens and the spectrum darkens as the grain size increases from 45\,$\mu$m to 500--1000\,$\mu$m (left panel of Figure~\ref{FoXX_Size}). Also, ``shoulder'' features (marked with arrows) at $\sim$0.8 and $\sim$1.3\,$\mu$m are symmetric around the main absorption at $\sim$1\,$\mu$m and flatten when decreasing the grain size. These shoulders are due to the overlapping $\alpha$, $\beta$ and $\gamma$ bands of olivine with $\gamma$ being 3-5 times stronger than $\alpha$ and $\beta$ \citep{shoulders}. NS2 has a deeper absorption and a more prominent shoulder at $\sim$1.3\,$\mu$m. So a smaller surface grain size on the surface of DH7 than in NS2 could explain the differences in the two spectra. 
The shallower 1\,$\mu$m absorption feature for Eureka could 
indicate the presence of yet smaller grains on its surface than the other two asteroids. 

\paragraph{Composition:} 
Changing the Fe to Mg ratio or Fo number of olivine affects the width of the 1\,$\mu$m feature as well as the reflectance. Smaller Fo numbers, i.e. higher content of Fe, have a wider 1\,$\mu$m feature and lower albedo (right panel of Figure~\ref{FoXX_Size}). The effect on the spectral profile is similar to that for particle size. In particular, the shoulders flatten when the Fo number increases. In our spectra they are non-symmetric around 1\,$\mu$m with the one from the longward side of the absorption being more prominent than from the shortward one. An alternative explanation for the differences in the spectra of the observed asteroids could therefore be that the olivine on the surface of NS2 is more Fe-rich than for DH47 and that Eureka has more Mg-rich olivine than the other asteroids. 

We conclude that the differences in the spectra between the Eureka family asteroids, if not due to systematics introduced during data acquisition and/or reduction, are caused by space weathering in combination with differences in surface particle size  and/or a different composition in terms of olivine Fo number. Because the shoulder from the longward side of the 1\,$\mu$m feature is more prominent in the spectra than the shoulder below 1\,$\mu$m, the dominant effect should be the difference in olivine composition, with NS2 more Mg-rich than DH47 and Eureka, and Eureka itself being more Fe-rich then the others.

\section{Conclusions and origin implications}
\paragraph{}Eureka family asteroids share a common olivine-dominated composition \citep{Borisov2017}, yet our analysis shows measurable differences between their spectra. We found these to be attributable to more than one causes: space weathering, particle grain size or composition. This last possibility would imply differences in the type of olivine between the three asteroids.

VF31 is the only L5 Trojan asteroid not part of the Eureka family. Its mineralogy is quite different from its companions; our analysis shows that it is spectrally similar to areas in the lunar highlands. The simplest spectral mixture that adequately represents VF31 features Mg-rich orthopyroxene and metallic iron as its principal ingredients. This interpretation is not unique and a better visual fit is achieved with mixtures dominated by iron metal and plagioclase with a small contribution by Mg-poor orthopyroxene.

Our study is based in part on observations by \citet{Rivkin2007}, therefore it is not surprising that this study's conclusions overlap with that work. Those authors placed VF31 within the pyroxene-rich S(VII) subclass of the \citet{Gaffey1993} classification scheme for S-type asteroids, a finding that led them to compare the asteroid spectrum with mixtures of iron-rich meteorite and primitive achondrite endmembers. Here we confirm the prevalence of iron metal and the existence of pyroxene. Identification of the latter is facilitated by the higher S/N of our visible spectrum while the former does a better job than our space weathering model at weakening the pyroxene absorption to match the observations. 
Additionally, we find that the best fits to the data are mixtures of orthopyroxene and meteoritic iron with or without plagioclase, rather than iron-rich primitive achondrites \citep{Rivkin2007}. The two interpretations are not mutually exclusive; the difference between \citeauthor{Rivkin2007} and this work is probably due to the increased availability of laboratory-measured endmember spectra for fitting since the previous study, rather than a fundamental difference between their approach and ours. Rapid progress in modelling the solar system's formation and early evolution also allows for a broader discussion on this asteroid's origin than was possible ten years ago. Below we examine the different types of candidate parent bodies and discuss the implications on a case-by-case basis, beginning with the proposed association with primitive achondrites, specifically the Acapulcoite-Lodranite (AL) meteorite group linked to S-type asteroids \citep{McCoy2000}. 

The provenance of AL meteorites was recently investigated by \citet{Neumann-2018_Icarus} using thermo-chronological modelling of potential parent bodies. They found the best-fit source objects to be large \mbox{(200--400\,km)}, partially differentiated planetesimals that formed relatively close to the Sun a few Myr after Calcium Aluminium-rich Inclusions, probably earlier and at a smaller heliocentric distance than the parent bodies of ordinary chondrites. These may still exist among the few tens of $>$200\,km Main Belt asteroids, a possibility also supported by the inferred shallow ($\sim$10\,km) excavation depth. Alternatively, they may have been shattered and collisionally dispersed during the early evolution of the terrestrial planets. 

If the presence of iron on the surface of VF31 does not reflect the original mineralogy of the parent body, it could represent extreme space weathering due to the asteroid's long residence time at 1.5\,AU from the Sun. Solar radiation intensity on the Martian Trojans is higher than for any Main Belt asteroid. Mars Crossers (MCs) and Near Earth Asteroids (NEAs), which come closer to the Sun, have typical dynamical lifetimes of $\lesssim 10^{8}$\,yr before being lost to collisions or dynamical ejection \citep{Bottke1994}, $\sim$40$\times$ less than the Trojans if the latter are primordial. Indeed, an NEA on a near-circular orbit resembling that of the Trojans would need to be located at $a \lesssim 1.5/\sqrt{40} \simeq 0.23$\,AU from the Sun to accumulate the same amount of incident solar energy over a period of $10^{8}$\,yr.

Our analysis shows a remarkable similarity between VF31 and the lunar surface. A quantitative assessment of the likelihood that VF31 could have originated from the Moon is beyond the scope of this work, yet it is reasonable to ask if this is at least possible. To answer this, we break the problem into the following sub-problems: i) Ejection  from the Moon ii) Transfer from the orbit of the Earth to an VF31-like orbit, and iii) Insertion into Mars's Trojan clouds.

An impact ejectum with a velocity higher than the Moon's escape velocity of 2.4\,km/s will leave its gravity field. To place this body in a heliocentric orbit, the ejection velocity should also exceed 3.5\,km/s \citep{Gladman1996}. According to \citet{Melosh1984}, a 1-km diameter object with this ejection velocity is produced by an impact with a projectile of incident speed of 10\,km/s and radius 125\,km to produce a 974\,km diameter crater \citep{Asphaug1997}. This is considerably smaller than the size of the largest lunar basin \citep[][D=2400\,km]{LunarBasinsList}, therefore this scenario is at least plausible.

Once the object is in heliocentric orbit, it must somehow find itself at an orbit similar to that of VF31, namely with a semi-major axis \mbox{a=1.52\,AU} and an inclination \mbox{i=31.5$^\circ$}. Here we consult the simulations by \citet{Gladman1995} who investigated the dynamical evolution of lunar impact ejecta. From their Figure 9, it indeed appears possible to reach such an orbit following escape from the Earth-Moon system.

The phase that relates to the last sub-question is potentially the most challenging as the object needs to be captured in stable libration around one of the Lagrange points of Mars. This scenario is  virtually impossible in the present era \citep{Polishook_Nature} but plausible during the final stages of terrestrial planet formation, when the orbit of Mars may have been perturbed by lunar-sized objects, resulting in a chaotic wandering of its semi-major axis \citep{Scholl2005}. In that case, the observed Martian Trojan asteroids were captured during one of the last semi-major axis jumps. A km-size fragment of the Moon could have been produced, for instance, during the Late Heavy Bombardment (LHB) $3.9$$\pm$$0.1$ Ga ago \citep{LHB1,LHB2} yet, if Mars formed 4.5\,Ga ago and its orbit stabilised in less than 200\,Myr \citep{Polishook_Nature}, a link between VF31 and the LHB appears unlikely. If we accept a lunar origin for the asteroid we must therefore conclude that it was ejected earlier, perhaps as soon as the lunar crust solidified \cite[eg][]{Marks_etal_JGRPlanets_2019}, or produced during the original cataclysmic event thought to have created our natural satellite \citep{HartmannDavis}. 

Alternatively, this asteroid may have originated from Mars itself. \citet{Polishook_Nature} recently proposed that the olivine-rich Mars Trojans at L5 originate from the martian mantle, having been excavated by an impact similar to that responsible for the hemisphere-wide Borealis basin \citep{Andrews-Hanna_Natur}. Initially deposited near the Martian orbit, the asteroids were captured permanently at L5 as Mars ``jumped'' to its present orbit in the final stages of its formation \citep{Scholl2005}. Although VF31 has higher orbital inclination than the olivine-rich Trojans, the dynamical component of this scenario will probably also work for that asteroid. However, the pyroxene-rich composition of (101429) is more compatible with crustal, rather than mantle, material \citep{Zuber_Nature}. In this context, our finding of strong iron metal contribution to the spectrum, if not related to space weathering, may be linked to the generally high abundance of iron oxides on the Martian surface \citep{Bibring-2005_Science}.

The Martian origin scenario is broadly consistent with our present understanding of Mars's surface and shallow interior composition, as a Borealis-scale impact would likely have ejected fragments of both crust and mantle into space. Pyroxene abundance on Mars's surface follows the planet's hemispherical dichotomy \citep{Bibring-2005_Science,Mustard-2005_Science,Riu-2019_Icarus}. Orthopyroxene (Low Calcium Pyroxene; LCP) is generally less abundant than clinopyroxene (High Calcium Pyroxene; HCP) and enriched in the older Noachian-era crust, while HCP is enriched in more recent lava flows, implying that the crust likely formed out of a fully melted magma where LCP and HCP were well mixed \citep{Bibring2009}. Our interpretation of the type of pyroxene present on the asteroid is somewhat at odds with the relative abundances of martian LCP and HCP; it could indicate depth-dependent survival of ejecta as intact chunks of material, with near-surface LCP-rich deposits more likely to survive than HCP-rich material deeper down; or simply a prevalence of LCP at the source depth on the time of ejection. The latter interpretation would be consistent with formation models of Noachian-era terrain \citep{Mustard-2005_Science} and suggests that the asteroid originated as a piece of Mars's ancient crust, dating from the earliest period of its existence.

Finally, we must also consider the possibility that VF31 is not a primordial occupant of the L5 cloud of Mars but was captured later, after the architecture of the solar system has settled into its
observed configuration. \cite{Christou2020} raised this possibility in relation to the close proximity of VF31's orbit to a secular resonance; their numerical simulations suggested that the Yarkovsky effect may have helped to stabilise the orbit post-capture. If that is the case, the most likely parent bodies of VF31 would then be extant or extinct S-type asteroids in the Main Belt (MB), probably its inner part. Dynamical transport from the MB to the Mars-crossing population tends to preserve the orbital inclination, therefore we favour a high-inclination source such as the Hungaria or Phocaea families \citep{Migliorini1998}. For instance, asteroids in the pyroxene-dominated Gaffey S(VII) subclass exist in the region occupied by the Hungarias, though with generally low abundance \citep{Lucas2016}.

\section{Prospects for Future Investigation}
We consider the present analysis of the L5 Martian Trojan asteroids and especially VF31 as the most detailed investigation of their mineralogy to-date. At the same time, we are aware of alternative approaches to mineralogical analysis of asteroid spectra, for instance through the Modified Gaussian Model \cite[MGM;][]{SunshinePieters1998,Sunshine.et.al2004} and we encourage their application on these data to obtain additional insight into these fascinating objects. Inevitably, however, making further progress in studying the Trojans will require the acquisition of new data.

The Large Synoptic Survey Telescope (LSST) should begin operations in 2022 and will eventually generate the largest catalog of Solar System objects to date - as faint as $r$=25  - as well as obtain {\it ugrizy} colours \citep{LSST}. It should find a few hundred additional members of the Eureka family \citep{Christou2020} and constrain the taxonomy of many of these new discoveries. Colours may be used to diagnose spectral differences among family members. From Figure~\ref{Sp3MarTr} we expect, for example, similar reflectance around 0.6\,$\mu$m but significant differences at 1.0\,$\mu$m, the central wavelengths of the {\it r} and {\it y} filters respectively. The projected 10 years of LSST nominal survey operation will span 5 different oppositions of the Mars Trojans, allowing also to characterise the spectral dependence on phase angle.

Higher quality visible and near-IR spectra on these faint asteroids could be obtained by the spectrograph \citep{HARMONI} on the planned European Extremely Large Telescope \cite[E-ELT;][]{EELT}. 
 Observations on yet longer wavelengths diagnostic of olivine composition have been previously obtained only for Eureka by the {\it Spitzer} space telescope \citep{Lim2011}. The soon-to-be-launched James Webb Space Telescope (JWST) willl bring much improved spectrophotomeric capabilities to asteroid studies in this part of the electromagnetic spectrum \citep{Rivkin2016}. Indeed, the superior sensitivity and spectral coverage of JWST allows the acquisition of high quality visible, near-IR and mid-IR spectra for much of the Eureka family as well as for (101429). This will allow more spectral comparisons of the type attempted here and over a wider spectral region.

Yet more comprehensive studies of these objects will likely require a spacecraft visit which could, en route to the Trojans, obtain spectra at Mars or the Moon for direct  comparison with the asteroid data \citep{Wickhusen2017}. 

\section{Acknowledgments}

\noindent We thank the two reviewers, David Polishook and Tomas Kohout, for their insightful comments that led to a considerably improved manuscript. 
This work was supported via grants (ST/M000834/1 and ST/R000573/1) from the UK Science and Technology Facilities Council. Astronomical research at the Armagh Observatory and Planetarium is grant-aided by the Northern Ireland Department for Communities.\\
Based on observations collected at the European Organisation for Astronomical Research in the Southern Hemisphere under ESO programme 296.C-5030 (PI: A.~Christou). \\
This research utilises spectra acquired with the NASA RELAB facility at Brown University.\\
Eureka spectral data utilised in this publication were obtained and made available by the MIT-UH-IRTF Joint Campaign for NEO Reconnaissance. The IRTF is operated by the University of Hawaii under Cooperative Agreement no. NCC 5-538 with the National Aeronautics and Space Administration, Office of Space Science, Planetary Astronomy Program. The MIT component of this work is supported by NASA grant 09-NEOO009-0001, and by the National Science Foundation under Grants Nos. 0506716 and 0907766.\\
This publication makes use of data products from the Near-Earth Object Wide-field Infrared Survey Explorer (NEOWISE), which is a project of the Jet Propulsion Laboratory/California Institute of Technology. NEOWISE is funded by the National Aeronautics and Space Administration.\\
We thank Andrew Rivkin for providing us with the SpeX NIR spectrum of (101429) 1998 VF31 and Boris Nedelchev for his suggestion to use Gauss-Hermite series polynomials in fitting the absorption features in the VF31 spectrum and discussions on the statistical significance of our mixture fits.

\appendix
\section{\label{sec:app1} Ranking of spectral matches}
In our spectral matching procedure (Section~\ref{match}), we assume the following simple error model:
\begin{itemize}
\item given a spectral measurement $x_i$ and its error $\sigma_i$, the real value $X_i$ is described by a Gaussian distribution with mean $x_i$ and standard deviation $\sigma_i$ (we assume uncorrelated measurement errors).
\item the distribution of the possible observed values is a Gaussian with mean $X_i$ and standard deviation $\sigma_i$, assuming that our experimental $\sigma_i$ are good estimates of the real error.
\end{itemize}

Note that the two assumptions above are equivalent to assuming that the distribution of the possible observed values is a Gaussian with mean $x_i$ and standard deviation $\sqrt{2}\sigma_i$, however the factor $\sqrt{2}$ has a small effect in the computations.

The cost functions $\chi^2_\mu$ and $\Phi_{std}$ cannot be reduced to a standard $\chi^2$ statistic. Therefore, we cannot take advantage of the known properties of the $\chi^2$ distribution and instead follow a numerical approach:
\begin{itemize} 
\item we generate a large number of clones of the observed spectrum substituting each measurement $x_i$ with a random quantity $x_i'$ generated from a parent Gaussian distribution with mean $x_i$ and standard deviation $\sqrt{2}\sigma_i$
\item each clone of the observed spectrum is normalised at the common normalisation wavelength
\item for each clone of the observed spectrum the cost functions are computed using $x_i'$ instead of $x_i$, a new ranking list is built, and the top-ranked model identified.
\end{itemize}

This approach allows to count the number of times a given model spectrum is ranked at the top of the score list and estimate the corresponding probability $P$ of being the top-ranked model.
\bibliographystyle{cas-model2-names}

\bibliography{MarsTrojanMineralogy}

\end{document}